\documentstyle[twocolumn,aps]{revtex}

\begin{document}
\draft

\twocolumn[\hsize\textwidth\columnwidth\hsize\csname @twocolumnfalse\endcsname

\title{Determining the density of states for  
classical statistical models: A random  walk algorithm 
to produce a flat histogram}

\author{Fugao Wang and D.  P. Landau}
\address{Center for Simulational Physics,
The University of Georgia, Athens, Georgia 30602} 
\date{\today}

\maketitle
\begin{abstract}               

We describe  an  efficient Monte Carlo algorithm using  a random walk in energy space 
to obtain a very accurate estimate of the density of states for  classical 
statistical models.
The density of states is modified at each step  
when the energy  level is visited to produce a flat histogram. 
By carefully   controlling the modification factor, we allow 
the density of states to converge to the true value very quickly, 
even for large systems. 
From the density of states at the  end of the random walk,
we can estimate thermodynamic quantities such as internal energy 
and specific heat capacity  
by calculating canonical averages at essentially any temperature.  
Using  this method, we not only can avoid repeating simulations 
at multiple temperatures, 
but can also estimate the Gibbs free energy and entropy, quantities 
which are not directly accessible  by  conventional Monte Carlo simulations. 
This  algorithm is especially  useful for complex systems 
with a rough landscape 
since all possible energy levels  are  visited 
with the same probability.  As with the multicanonical Monte Carlo  
technique, our method  overcomes the tunneling barrier 
between coexisting  phases at first-order phase transitions. 
In this paper,  we apply our   algorithm  to both 1st 
and 2nd  order phase transitions to 
demonstrate  its efficiency and accuracy.
We obtained direct simulational estimates for the density of states for 
two-dimensional ten-state Potts models  on lattices up to $200 \times 200 $
and Ising models on lattices up to  $256 \times 256$.
Our simulational results  are  compared to both
exact solutions and existing  numerical data obtained using other methods.
Applying this approach to a 3D $\pm J$  spin glass model we
estimate the  internal energy and entropy  at zero temperature; and, 
using a two-dimensional random walk in energy and order-parameter space, 
we obtain the (rough) canonical distribution and  energy 
landscape in order-parameter space.  Preliminary data suggest that the 
glass transition temperature is about $1.2$ and that better estimates
can be obtained with more extensive application of the method.  
This simulational method is not restricted to energy space  
and can be used to calculate the density of states  for any parameter 
by a random walk in the corresponding  space.  
\end{abstract}


]

\section{Introduction}

Computer simulation now plays a major role in statistical 
physics~\cite{landau_binder}, particularly for the study 
of phase transitions and
critical phenomena. The standard Monte Carlo (MC)  
method has been the Metropolis
importance sampling algorithm~\cite{metropolis}, but more recently new,
efficient algorithms have begun to play a role in allowing simulation to
achieve the resolution which is needed to accurately locate and characterize
phase transitions~\cite{landau_binder}. For example, cluster flip
algorithms, beginning with the seminal work of 
Swendsen and Wang~\cite{swendsen_wang} 
and extended by Wolff~\cite{wolff}, have been used to reduce
critical slowing down near 2nd order transitions. Similarly, the
multicanonical ensemble method~\cite{berg_0} was introduced to overcome the
tunneling barrier between coexisting phases at 1st order transitions and has
general utility for systems with a rough energy 
landscape~\cite{janke_kappler,berg_1,alves}. 
In both situations, histogram re-weighting
techniques~\cite{ferrenberg} can be applied in the analysis to increase the
amount of information that can be gleaned from simulational data, but the
applicability of re-weighting is severely limited in large systems by the
statistical quality of the ``wings'' of the histogram. This latter effect is
quite important in systems with competing interactions for which short range
order effects might occur over very broad temperature ranges or even give
rise to frustration that produces a very complicated energy landscape and
limits the efficiency of ``standard'' methods.

One of the most important quantities in statistical physics is the density
of states $g(E)$, i.e. the number of all possible states (or configurations)
for an energy level $E$ of the system, but direct estimation of this
quantity has not been the goal of simulations. Instead, most conventional
Monte Carlo algorithms~\cite{landau_binder} such as Metropolis importance
sampling, Swendsen-Wang cluster flipping, etc. generate a canonical
distribution $g(E)e^{-E/k_{B}T}$ at a given temperature. Such distributions
are so narrow that, with conventional Monte Carlo simulations, multiple runs
are required if we want to know thermodynamic quantities over a significant
range of temperatures. In the canonical distribution, the density of states
does not depend on the temperature at all. If we can estimate the density of
states $g(E)$ with high accuracy for all energies, we can then construct
canonical distributions at essentially any temperature. For a given model in
statistical physics, once the density of states is known we can calculate
the partition function as $Z=\sum\limits_{E }g(E)e^{-\beta E}$, and the
model is essentially ``solved" since most thermodynamic quantities can be
calculated from it. Though computer simulation is already a very powerful
method in statistical physics~\cite{landau_binder}, it seems that there is
no efficient algorithm to calculate the density of states very accurately
for large systems. Even for exactly solvable models such as the 2-dim Ising
model, $g(E)$ is impossible to calculate exactly 
for a large system~\cite{beale}.

The multicanonical ensemble method~\cite{berg_0,berg_00} proposed by Berg 
{\it et al.} has been proved to be very efficient in studying first-order
phase transitions where simple canonical simulations have difficulty
overcoming the tunneling barrier between coexisting phases at the transition
temperature~\cite{berg_0,janke_kappler,janke_4,berg_4,janke_3}. The method
also has been successfully applied to some complex systems, such as spin
glass models~\cite{berg_1,janke_3,berg_2,berg_3,berg_6,hatano} and protein
folding problems~\cite{alves}, for which the energy landscape is very rough
and the conventional canonical Monte Carlo simulation gets easily trapped in
local minima at low temperature. In the multicanonical method, we have to
estimate the density of states $g(E)$ first, then perform a random walk with
a probability of ${\frac 1 {g(E)}}$ to make the histogram flat in the
desired region in the phase space, such as between two peaks of the
canonical distribution at the first-order transition temperature. In a
multicanonical simulation, the density of states need not necessarily be
very accurate, as long as the simulation generates a relatively flat
histogram and overcomes the tunneling barrier in energy space. This is
because the subsequent re-weighting \cite{berg_0,berg_00}, does not depend
on the accuracy of the density of the states as long as the histogram can
cover all important energy levels with sufficient statistics. (If the
density of states could be calculated very accurately, then the problem
would have been solved in the first place and we needn't perform any further
simulation such as with the multicanonical simulational method.) Berg {\it 
et al.} proposed a recursive method to calculate the density of states by
accumulating histogram entries and estimating the density of states
iteratively with the histogram data~\cite{berg_1,berg_7}. Their method works
well for small systems since the number of all possible states is small;
however, for large systems the number of possible states increases
exponentially with the size of a system. A simple iteration method to
construct a histogram gets trapped in a narrow energy range, and it needs an
extremely long time and large number of iterations to get an accurate
estimate for the density of states for the entire energy space, even for a
small system. It is thus not practical to calculate the density of states
for large systems with this approach. For a simple model such as the Potts
model, Berg and Neuhaus used finite-size scaling theory to ``guess" the
density of states up to $100 \times 100 $ from the simulational results of
the small systems~\cite{berg_1}. Such results are not as reliable as direct
simulations, moreover for complex systems such as spin glass models, we can
not simply apply such finite-size scaling at all. Berg and Celik applied
their multicanonical method to a 2D spin-glass model, but they could test
their method only up to a $48 \times 48$ lattice~\cite{berg_1}. Recently
Berg and Janke proposed a similar method (multioverlap simulational) for the 3D
Ising spin-glass model, they obtained some reliable results for systems as
large as $12 \times 12 \times 12$~\cite{berg_2}.

Lee~\cite{Lee} independently proposed the entropic sampling method, which is
basically equivalent to multicanonical ensemble sampling. He used an
iteration process to calculate the microcanonical entropy at $E$ which is
defined by $S(E)=\ln[g(E)]$ where $g(E)$ is the density of states. He also
applied his method to the 2D ten-state ($Q=10$) Potts model and the 3D Ising
model; however, just as for other simple iteration methods, it works well
only for small systems. He obtained a good result with his method for the $
24 \times 24$ 2-dim $Q=10$ Potts model and the $4 \times 4 \times 4 $ 3-dim
Ising model.

Oliveira {\it et al.}~\cite{oliveira_0,oliveira_1,oliveira_2}
proposed the broad histogram method with which they calculated the density
of states by estimating the probabilities of possible transitions between
all possible states of a random walk in energy space. Using simple canonical
average formulae in statistical physics, they then calculated thermodynamic
quantities for any temperature. Though the authors believed that the broad
histogram relation is exact, their simulational results have systematic
errors even for the Ising model on a lattice as small as $32 \times 32$ in
references~\cite{oliveira_0,wang_0}. They believed that the error was due to
the particular dynamics adopted within the broad histogram 
method~\cite{oliveira_4}, but other work~\cite{wang_2,berg_5} argued that it violates
the detailed balance condition. The algorithm was corrected in 
reference~\cite{wang_3} and an approach to 
the broad-histogram method was proposed to
calculate the density of states based on the number of potential moves
during the random walk in energy space~\cite{wang_0}. They obtained more
accurate results for a $32 \times 32 $ Ising model than the broad histogram
did; but this method also suffers from the systematic errors and substantial
deviations when system becomes larger than $32 \times 32$~\cite{wang_0,Lima}.

It is thus an extremely difficult task to calculate density of states
directly with high accuracy for large systems. All methods based on
accumulation of histogram entries, such as the histogram method of
Ferrenberg and Swendsen~\cite{ferrenberg}, Lee's version of
multicanonical method (entropic sampling)~\cite{Lee}, broad histogram 
method~\cite{oliveira_0,wang_0,Lima} and 
flat histogram method~\cite{wang_0} have
the problem of scalability for large systems. These methods suffer from
systematic errors when systems are large, so we still need a superior
algorithm to calculate the density of states for large systems.

Very recently, we introduced a new, general, efficient Monte Carlo algorithm
that offers substantial advantages over existing 
approaches~\cite{wang_landau}. In this paper, 
we will explain the algorithm in detail,
including our implementation, and describe its application not only to 1st
and 2nd order phase transitions, but also to a 3D spin glass model that has
a rough energy landscape.

Unlike conventional Monte Carlo methods that directly generate a canonical
distribution at a given temperature $g(E)e^{-E/K_{\text B}T}$, our approach
is to directly estimate the density of states $g(E)$ accurately via a random
walk that produces a flat histogram in energy space. We modify density of
states at each step of the random walk, and by carefully controlling the
modification factor we can obtain a density of states that converges to the
real value very quickly even for large systems. The resultant density of
states is accurate enough to calculate thermodynamic quantities by applying
canonical average formulas in statistical physics.

The remainder of this paper is arranged as follows. In section II, we
present our general algorithm in detail. In section III, we apply our method
to the 2D $Q=10$ Potts model which has a first-order phase transition. In
section IV, we apply our method to a model with a second-order phase
transition to test the accuracy of the algorithm. In section V, we consider
the 3D $\pm J$ spin glass model, a system with rough landscapes. Discussion
and the conclusion are presented in section VI.

\section{A general and efficient algorithm to estimate the density of states
with a flat histogram}

Our algorithm is based on the observation that if we perform a random walk
in energy space by flipping spins randomly for a spin system, and the
probability to visit a given energy level $E$ is proportional to the
reciprocal of the density of states ${\frac 1 {g(E)}}$, then a flat
histogram is generated for the energy distribution. This is accomplished by
modifying the estimated density of states in a systematic way to produce a
``flat'' histogram over the allowed range of energy and simultaneously
making the density of states converge to the true value. We modify the
density of states constantly during each step of the random walk and use the
updated density of states to perform a further random walk in energy space.
The modification factor of the density of states is controlled carefully,
and at the end of simulation the modification factor should be very close to 
$1$ which is the ideal case of the random walk with the true density of
states.

At the very beginning of our simulation, the density of states is {\it a
priori} unknown, so we simply set all entries to $g(E)=1$ for all possible
energies $E$. Then we begin our random walk in energy space by flipping
spins randomly and the probability at a given energy level is proportional
to ${\frac 1 {g(E)}}$. In general, if $E_1$ and $E_2$ are energies before
and after a spin is flipped, the transition probability from energy level 
$E_1$ to $E_2$ is: 
\begin{equation}
p(E_1 \rightarrow E_2)= \min\left[{\frac {g(E_1)} {g(E_2)}}, 1\right].  \label{eqn:p}
\end{equation}
Each time an energy level $E$ is visited, we modify the existing 
density of states  by a
modification factor $f>1$, i.e. $g(E) \rightarrow g(E)*f$. (In practice, 
we use the formula 
$\ln[g(E)] \rightarrow \ln[g(E)]+\ln(f)$ in order to fit all possible
$g(E)$ into double precision numbers for the systems we will discuss in this
paper.) If the random walk rejects a possible move and stays at the
same energy level, we also modify the existing density of states with the
same modification factor. Throughout this study we have used an initial
modification factor of $f=f_0=e^1\simeq2.71828...$ which allows us to reach
all possible energy levels very quickly even for a very large system. If $f_0
$ is too small, the random walk will spend an extremely long time to reach
all possible energies. However, too large a choice of $f_0$ will lead to
large statistical errors. In our simulations, the histograms are generally
checked about each 10000 MC sweeps. A reasonable choice is to make 
$f_0^{10000}$ have the same order of magnitude as the total number of states
($Q^N$ for a Potts model). During the random walk, we also accumulate the
histogram $H(E)$ (the number of visits at each energy level $E$) in the
energy space. When the histogram is ``flat" in the energy range of the
random walk, we know that the density of states converges to the true value
with an accuracy proportional to that modification factor $\ln(f)$. Then we
reduce the modification factor to a finer one using a function like $f_1=
\sqrt{f_0}$, reset the histogram, and begin the next level random walk
during which we modify the density of states with a finer modification
factor $f_1$ during each step. We continue doing so until the histogram is
``flat" again and then reduce the modification factor $f_{i+1}=\sqrt{f_i}$
and restart. We stop the random walk when the modification factor is smaller
than a predefined value (such as $f_{\text{final}}=\exp(10^{-8})\simeq
1.00000001$). It is very clear that the modification factor acts as a
most important control parameter for the accuracy of 
the density of states during the
simulation and also determines how many MC sweeps are necessary for the
whole simulation.
The accuracy of the density of states  depends 
on  not only  $f_{\text {final}}$  but also  many other factors, 
such as  the complexity and size of the system, 
criterion of the flat histogram  and other details of the implementation 
of the algorithm. 

It is impossible to obtain a perfectly flat histogram and
the phrase ``flat histogram'' in this paper means that histogram $H(E)$ for
all possible $E$ is not less than $x\%$ of the average histogram $\langle
H(E)\rangle $, where $x\%$ is chosen according to the size and complexity of
system and the desired accuracy of the density of states. For the $L=32$, 2D
Ising model with only nearest-neighbor couplings, this percentage can be
chosen as high as $95\%$, but for large systems the criterion for
``flatness" may never be satisfied if we choose too high a percentage and
the program may run forever.

One essential constraint on the implementation of the algorithm is that 
the density of states during the random walk converges to the true value. 
The algorithm  proposed in this paper has this property. The
accuracy of the density of states is proportional to $\ln(f)$ at that
iteration; however, $\ln(f_{\text {final}})$ can not be chosen arbitrary
small or the modified $\ln[g(E)]$ will not differ from the unmodified one
to within the number of digits in the double precision numbers used in the
calculation. If this happens, the
algorithm no longer converges to the true value, and the program may run
forever. Even if  $f_{\text {final}}$ is within range but too small, 
the calculation
might take excessively long to finish. 

We have chosen to reduce the modification factor by a square root function,
and $f$ approaches $1$ as the number of iterations approaches infinity. In
fact, any function may be used as long as it decreases $f$ monotonically to 
1. A simple and efficient formula is $f_{i+1}=f_{i}^{1/n}$, where $n>1$.
The value of $n$ can be chosen according to the available CPU time and
expected accuracy of the simulation. For the systems that we have studied
the choice of $n=2$ yielded good accuracy in a relatively short time, even
for large systems. When the modification factor is almost $1$ and the random
walk generates a uniform distribution in energy space, the density of states
should converge to the true value for the system.

The method can be further enhanced by performing multiple random walks, each
for a different range of energy, either serially or in parallel fashion. We
restrict the random walk to remain in the range by rejecting any move out of
that range. The resultant pieces of the density of states can then be joined
together and used to produce canonical averages for the calculation of
thermodynamic quantities at essentially any temperature.

We should point out here that during the random walk (especially at the
early stage of iteration) in the energy space, the algorithm does not
satisfy the detailed balance condition exactly, since the density of states
is modified constantly during the random walk. However, after many
iterations, the density of states converges to the true value very quickly
as the modification factor approaches $1$. If $p(E_1 \rightarrow E_2)$ is
the transition probability from the energy level $E_1$ to level $E_2$, from
equation (\ref{eqn:p}), the ratio of the transition probabilities from $E_1$
to $E_2$ and from $E_2$ to $E_1$ can be calculated very easily as: 
\begin{equation}
{\frac {p(E_1 \rightarrow E_2)} {p(E_2 \rightarrow E_1)}} = {\frac {g(E_1)}{
g(E_2)}}
\end{equation}
where $g(E)$ is the density of states. In another words, our random walk
algorithm satisfies the detailed balance condition: 
\begin{equation}
{\frac {1} {g(E_1)} }p(E_1 \rightarrow E_2) ={\frac {1} {g(E_2)} }{p(E_2
\rightarrow E_1)}
\end{equation}
where ${\frac 1 {g(E_1)} }$ is the probability at the energy level $E_1$ and 
$p(E_1 \rightarrow E_2)$ is the transition probability from $E_1$ to $E_2$
for the random walk. We conclude that the detailed balance condition is
satisfied with accuracy proportional to the modification factor $\ln(f)$.

Almost all recursive methods update the density of states by using the
histogram data directly only after enough histogram entries are 
accumulated~\cite{berg_0,berg_1,berg_4,berg_2,berg_3,berg_6,hatano,berg_5,hansmann_1,hansmann_2,janke_2}. 
Because of the exponential growth of the density of states in energy
space, this process is not efficient because the histogram is accumulated
linearly. In our algorithm, we modify the density of states at each step of
the random walk, and this allows us to approach the true density of states
much faster than conventional methods especially for large systems. (We also
accumulate histogram entries during the random walk, but we only use it to
check whether the histogram is flat enough to go to the next level random
walk with a finer modification factor.)

We should point out here that the total number of configurations increases
exponentially with the size of the system; however, the total number of
possible energy levels increases linearly with the size of system. It is
thus easy to calculate the density of states with a random walk in energy
space for a large system. In this paper for an example, we consider the
Potts model on a $L\times L$ lattice with nearest-neighbor 
interactions~\cite{fywu}. 
For $Q \geq 3$, the number of possible energy levels is about $2N$,
where $N=L^2$ is the total number of the lattice site. However, the average
number of possible states (or configurations) on each energy level is as
large as ${\frac {Q^N} {2N}}$, where $Q$ is the number of possible states of
a Potts spin and $Q^N$ is the total number of possible configurations of the
system. This is the reason why most models in statistical physics are well
defined, but we can not simply use our computers to realize all possible
states to calculate any thermodynamic quantities, this is also the reason
why efficient and fast simulational algorithms are required in the numerical
investigations.

By the end of simulation, we only obtain relative density, since the density
of states can be modified at each time it is visited. We can apply the
condition that the total number of possible states for the $Q$ state Potts
model is $\sum\limits_{E} g(E)=Q^N$ or the number of ground state
 is $Q$ to get the absolute density of states.

\section{Application to a first order phase transition}

\subsection{Potts model and its canonical distribution}

In this section, we apply our algorithm to a model with a first-order phase
transition~\cite{binder_landau_1,landau_potts}. We choose the 2D ten state
Potts model since it serves as an ideal laboratory for temperature-driven
first-order phase transitions. Since some exact solutions and extensive
simulational data are available, we have ample opportunity to compare our
results with other values.

\begin{figure}[h]
\caption{ Density of states $g(E)$ for the 2-dim $Q=10$ Potts model as a
function of energy per lattice site $E/N$. With the scale in the figure, the
errors of the simulational data are within the width of the lines. }
\label{fig:potts_density}
\includegraphics{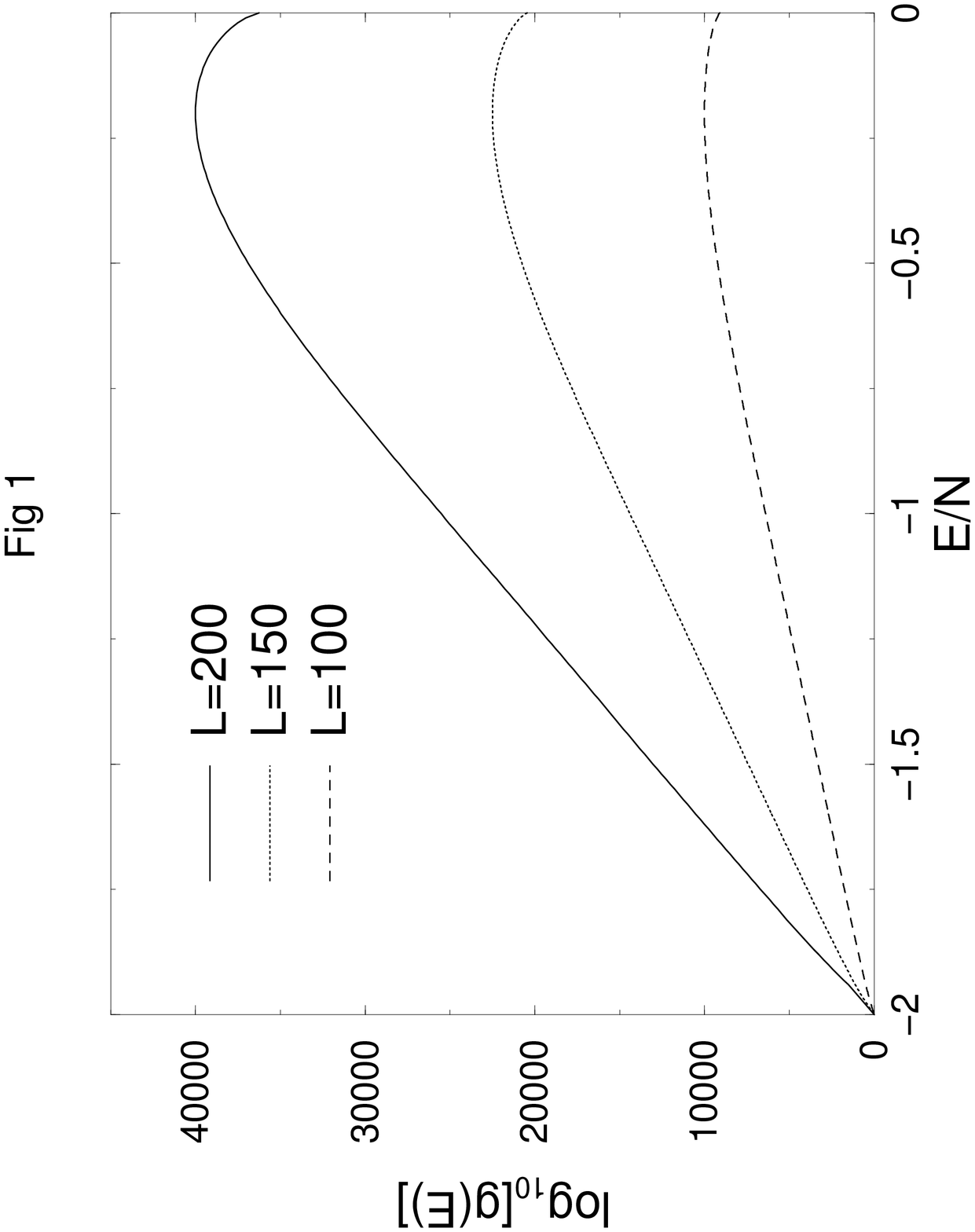}\vspace{3.0in}
\end{figure}

We consider the 2-dimensional $Q=10$ Potts model on $L \times L$ square
lattice with nearest-neighbor interactions and periodic boundary conditions.
The Hamiltonian for this model can be written as: 
\begin{equation}
{\cal H}=-\sum\limits_{<ij>}\delta(q_i, q_j)
\end{equation}
and $q=1,2,...Q$. During the simulation, we select lattice sites randomly
and choose integers between $[1:Q]$ randomly for new Potts spin values. The
modification factor $f_i$ changes from $f_0=e^{1}=2.71828$ at the very
beginning to $f_{\text{final}}=\exp(10^{-8})\simeq 1.00000001$ by the end of
the random walk. At the end of the simulations, our algorithm only provides
a relative density of states for different energies, so to extract the
correct density of states, we can either use the fact that the total number
of possible states is $Q^N$ or that the number of ground states is $Q$,
where $N=L^2$ is the total number of lattice sites. (Actually we can use one
of these two conditions to get the absolute density of states, and use the
other condition to check the accuracy of our result.) To guarantee the
accuracy of thermodynamic quantities at low temperatures in further
calculations, in this paper we use the condition that the number of the
ground states is $Q$ to normalize the density of states. The densities of
states for $100 \times 100$, $150 \times 150 $ and $200 \times 200 $ are
shown in Fig.~\ref{fig:potts_density}. It is very clear from the figure that
the maximum density of states for $L = 200 $ is very close to $10^{40000}$
which is actually about $5.75 \times 10^{39997}$ from our simulational data.

\begin{figure}[h]
\caption{ The canonical distributions at the transition temperature $P(E,T_{%
\text c})=g(E)e^{-E/K_BT_{\text{c}}}$ for the $Q=10$ Potts model for (a) $%
L=60$, 80 and 100 (single random walk) and (b) $L=150$ and 200 (multiple
random walks). The insets show the histograms of the random walks to
estimate the densities of states. }
\label{fig:potts_can_1}

(a) \\
\includegraphics{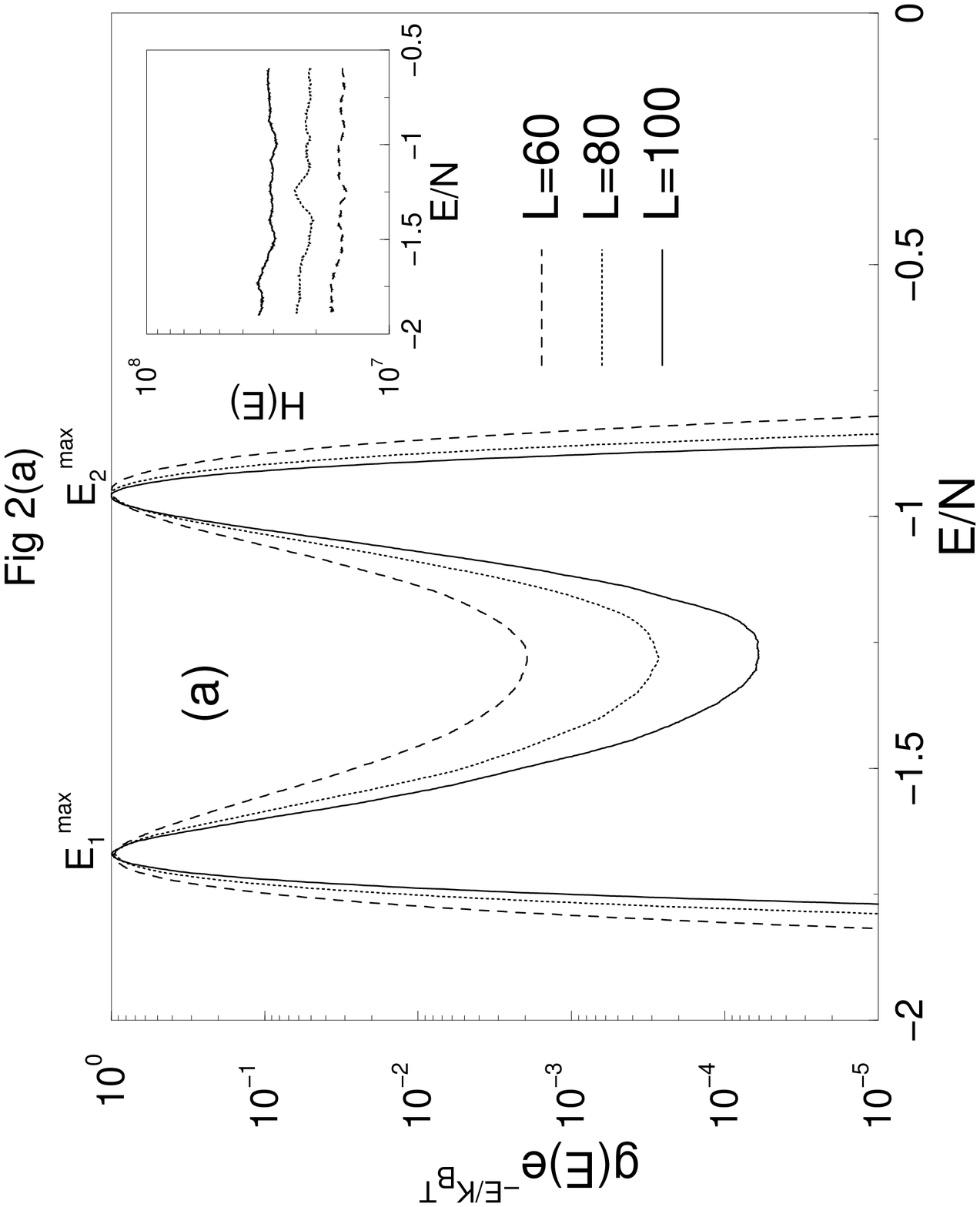}
\vspace{3.0in}
\newpage

(b)  \\
\includegraphics{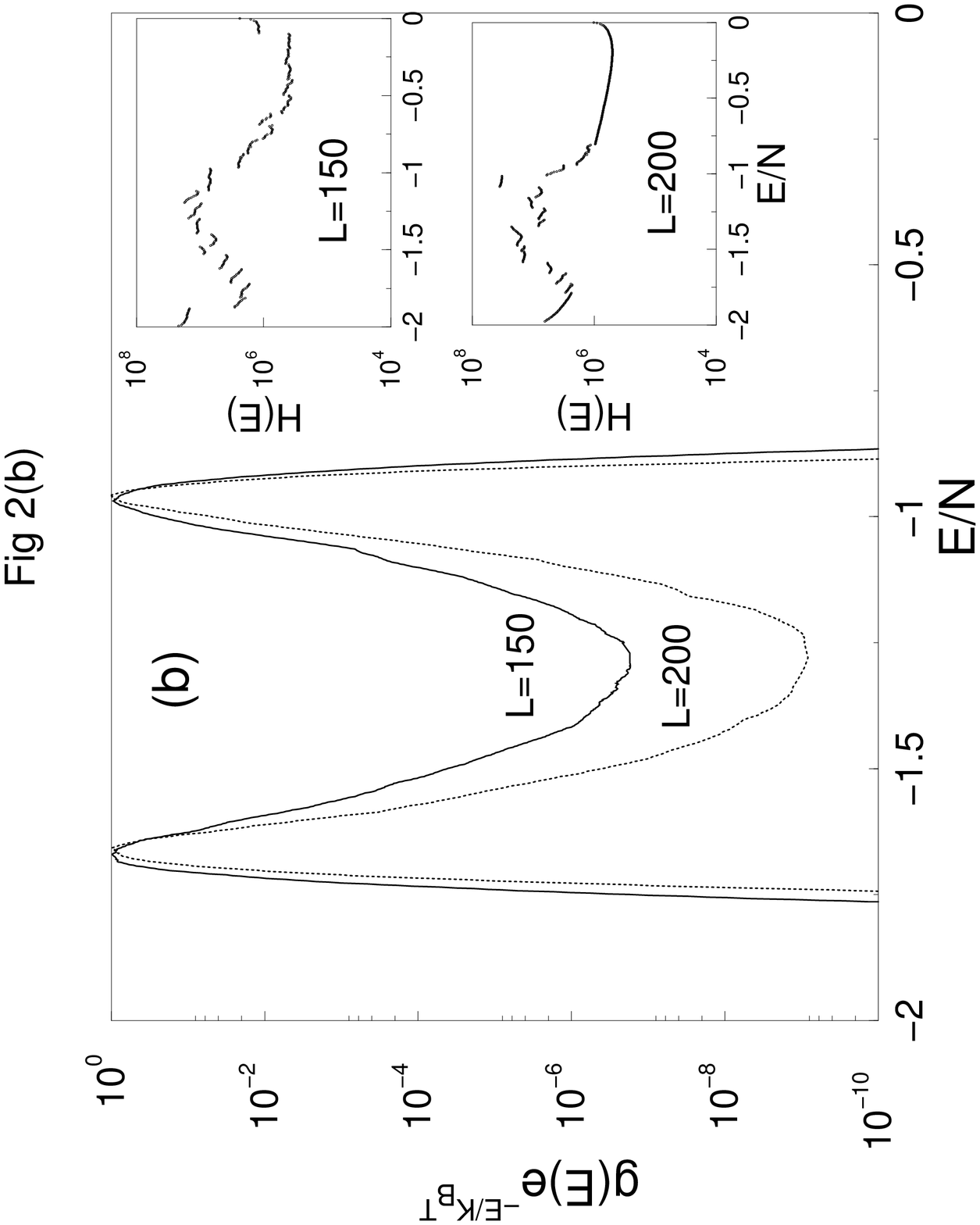}\vspace{3.0in}

\end{figure}

Conventional Monte Carlo simulation (such as Metropolis 
sampling~\cite{landau_binder,metropolis}) realizes 
a canonical distribution $P(E, T)$ by
generating a random walk Markov chain at a given temperature: 
\begin{equation}
P(E, T)=g(E)e^{-E/k_BT}
\end{equation}
\noindent From the simulational result for the density of states $g(E)$, we
can calculate the canonical distribution by the above formula at essentially
any temperature without performing multiple simulations. 
In Fig.  \ref{fig:potts_can_1}(a), we show the resultant double peaked canonical
distribution~\cite{landau_potts}, at the transition temperature $T_{\text{c}}$ 
for the first-order transition of the $Q=10$ Potts model. The
``transition temperatures" $T_{\text{c}}(L)$ are 0.70171 for $L=60$, 0.70143
for $L=80$ and 0.70135 for $L=100$ which are determined by the temperatures
where the double peaks are of the same height. Note that the peaks of the
distributions are normalized to 1 in this figure. The valley between two
peaks is quite deep. e.g. is $7 \times 10^{-5}$ for $L=100$. The latent heat
for this temperature driven first-order phase transition can be estimated
from the energy difference between the double peaks. Our results for the
locations of the peaks are listed in the Table 1. They are consistent with
the results obtained by multicanonical method~\cite{berg_00} and multibondic
cluster algorithm~\cite{janke_kappler} for those lattice sizes for which
these other methods are able to generate estimates. As the table shows, our
method produces results for substantially larger systems than have been
studied by these other approaches.

Because of the double peak structure at a first-order phase transition,
conventional Monte Carlo simulations are not efficient since an extremely
long time is required for the system to travel from one peak to the other in
energy space. With the algorithm proposed in this paper, all possible energy
levels are visited with equal probability, so it overcomes the tunneling
barrier between the coexisting phases in the conventional Monte Carlo
simulations. The histograms for $L$ = 60, 80 and 100 are shown in the inset
of the Fig.~\ref{fig:potts_can_1}(a) and  are very flat. The histogram in
the figure is the overall histogram defined by the total number of visits to
each energy level for the random walk. Here, too, we choose the initial
modification factor $f_0 = e^1$, and the final one as $\exp(10^{-8})\simeq
1.00000001$; and the total number of iterations is 27. In our simulation, we
do not set a predetermined number of MC sweeps for each iteration but rather
give the criterion which the program checks periodically. Generally, the
number of MC sweeps needed to satisfy the criterion increases as we reduce
the modification factor to a finer one, but we cannot predict the exact
number MC sweeps needed for each iteration before the simulation. We believe
that it is preferable to allow the program to decide how great a
simulational effort is needed for a given modification factor $f_i$. This
also guarantees a sufficiently flat histogram resulting from a random walk
which in turn determines the accuracy of the density of states at the end of
the simulation. We nonetheless need to perform some test runs to make sure
that the program will finish within a given time. The entire simulational
effort used was about $1.6 \times 10^7$ visits per energy level (or 
$3.2\times 10^7$ MC sweeps) for $L=60$, $2.2 \times 10^7$ visits for $L=80$
and $3.3 \times 10^7$ visits for $L=100$. With the program we implemented,
the simulation for $L=100$ can be completed within two weeks in a single 
600 MHz Pentium III processor.

To speed up the simulation, we needn't constrain ourselves to performing a
single random walk over the entire energy range with high accuracy. If we
are only interested in a specific temperature range, such as near $T_{\text{c
}}$, we could first perform a low precision unrestricted random walk, i.e.
over all energies, to estimate the required energy range, and then carry out
a very accurate random walk for the corresponding energy region. The inset
of Fig.~\ref{fig:potts_can_1}(a) only shows the histograms for the extensive
random walks in the energy range between $E/N=-1.90$ and $-0.6$. If we need
to know the density of states more accurately for some energies, we also can
perform separate simulations, one for low energy levels, one for high energy
levels, the other for middle energy which includes double peaks of the
canonical distribution at $T_{\text{c}}$. This scheme not only speeds up the
simulation, but also increases the probability of accessing the energy
levels for which both maximum and minimum values of the distributions occur
by performing the random walk in a relatively small energy range. If we
perform single random walk over all possible energies, it will take a long
time to generate rare spin configurations. Such rare energy levels include
the ground energy level or low energy levels with only few spins with
different values and high energy levels where all, or most, adjacent Potts
spins have different values.

With the algorithm in this paper, if the system is not larger than $100
\times 100$, the random walk on important energy regions (such as that which
includes the two peaks of the canonical distribution at $T_{\text{c}}$) can
be carried out with a single processor and  will give an accurate density of
states within about $10^{7}$ visits per energy level. The results shown in
Fig.~\ref{fig:potts_can_1}(a) were obtained using a single processor.
However, for a larger system, we can use a parallelized algorithm by
performing random walks in different energy regions, each using a different
processor. We have implemented this approach using PVM with a simple
master-slave model and can then obtain an accurate estimate for the density
of states with relatively short runs on each processor. The densities of
states for $150 \times 150$ and $200 \times 200 $, shown in Fig.~\ref%
{fig:potts_density}, were obtained by joining together the estimates
obtained from $21$ independent random walks, each constrained within a
different regions of energy. The histograms from the individual random walks
are shown in the inset of Fig.~\ref{fig:potts_can_1}(b) both for $150 \times
150 $ and $200 \times 200$ lattices. In this case, we only require that the
histogram of the random walk in the corresponding energy segment is
sufficiently flat without regard to the relative flatness over the entire
energy range. In Fig.~\ref{fig:potts_can_1}(b), the results for large
lattices show clear double peaks for the canonical distributions at
temperatures $T_{\text{c}}(L)=0.70127$ for $L=150$ and $T_{\text{c}%
}(L)=0.701243$ for $L=200$. The exact result is $T_{\text{c}}=0.701232....$
for the infinite system. Considering the valley which we find for $L=200$ is
as deep as $9\times 10^{-10}$, we can understand why it is impossible for
conventional Monte Carlo algorithms to overcome the tunneling barrier with
available computational resources.

If we compare the histogram for $L=60$ in Fig.~\ref{fig:potts_can_1}(a) with
that for $L=200 $ in Fig.~\ref{fig:potts_can_1}(b), we see very clearly that
the simulation effort for $L=200 $ ($9.8 \times 10^6$ visits per energy
level) is even less then the effort for $L=60$ ($1.6 \times 10^7$ visits per
energy level.). It is more efficient to perform random walks in relatively
small energy segments than a single random walk over all energies. The
reason is very simple, the random walk is a local walk, which means for a
given $E_1$, the energy level for the next step only can be one of 9 levels
in the energy range $[E_1-4, E_1+4]$. (for the Potts model discussed in this
section). The algorithm itself only requires that the histogram on such
local transitions is flat. (A single random walk, subject to the requirement
of a flat histogram for all energy levels, will take quite long.) For random
walks in small energy segments, we should be very careful to make sure that
all spin configurations with energies in the desired range can be equally
accessed so we restart the random walk periodically from independent spin
configurations.

An important question that must be addressed is the ultimate accuracy of the
algorithm. One simple check is to estimate the transition temperature of the
2D $Q=10$ Potts model for $L=\infty$ since the exact solution is known.
According to finite-size scaling theory, the ``effective" transition
temperature for finite systems behaves as: 
\begin{equation}
T_{\text{c}}(L)=T_{\text{c}}(\infty)+{\frac c {L^d}}
\end{equation}
where $T_{\text{c}}(L)$ and $T_{\text{c}}(\infty)$ are the transition
temperatures for finite- and infinite-size systems, respectively, $L$ is the
linear size of the system and $d$ is dimension of the lattice.

\begin{figure}[h]
\caption{ Extrapolation of finite lattice ``transition temperatures" for the 
$Q=10$ Potts model. }
\label{fig:potts_tc}

\includegraphics{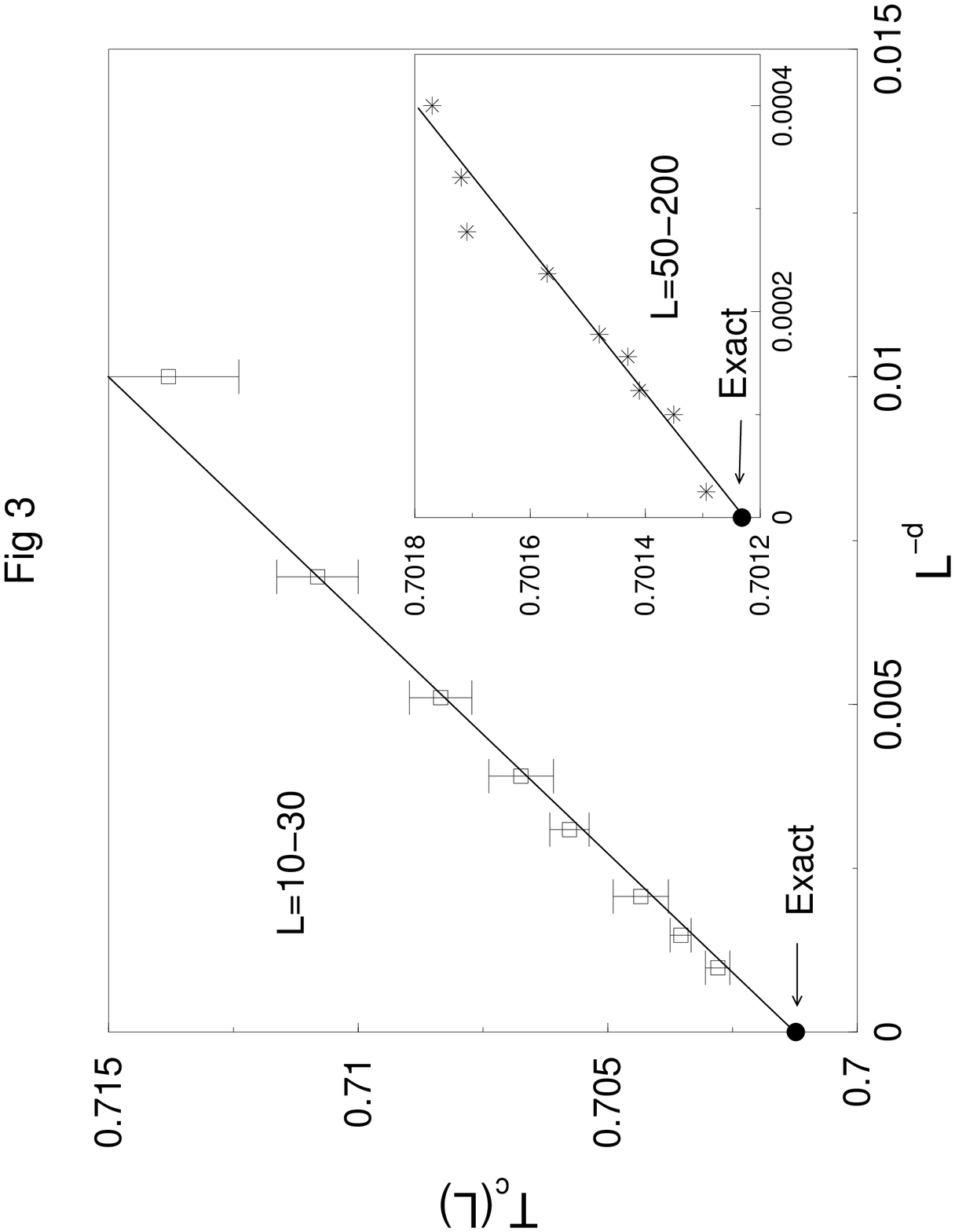}
\vspace{3.0in}

\end{figure}

In Fig.~\ref{fig:potts_tc}, the transition temperature is plotted as a
function of $L^{-d}$. The data in the main portion of the figure are
obtained from small systems($L=10\sim30$), and the error bars are estimated
by results from multiple independent runs. 
Clearly the transition
temperature extrapolated from our simulational data is 
$T_{\text c}(\infty)=0.7014 \pm 0.0004$ which is 
consistent with the
exact solution 
($T_{\text{c}}=0.701232....$)
for the infinite system. 
To get an even more accurate
estimate, and also test the accuracy of the density of states from single
runs for large systems, we performed single, long random walk on large
lattices($L=50\sim200$). The results, plotted as a function of lattice size
in the inset of the figure, show that the transition temperature
extrapolated from the finite systems is $T_{\text c}(\infty)=0.701236 \pm 0.000025$
which is still consistent with the exact
solution. 

We also compare our simulational result for the $Q=10$ Potts model with the
existing numerical data such as estimates of transition temperatures and
double peak locations obtained with the multicanonical simulational method
by Berg and Neuhaus~\cite{berg_0} and the Multibondic cluster algorithm by
Janke and Kappler~\cite{janke_kappler}. All results are shown in Table 1.
With our random walk simulational algorithm, we can calculate the density of
states up to $200 \times 200 $ within $10^7$ visits per energy level to
obtain a good estimate of the transition temperature and locations of the
double peaks. Using the multicanonical method and a finite scaling guess for
the density of states, Berg {\it et al.} only obtained results for lattices
as large as $100 \times 100$~\cite{berg_0}, and multibondic cluster
algorithm data~\cite{janke_kappler} were not given for systems larger than 
$50 \times 50$.

In section IV, the accuracy of our algorithm will be further tested by
comparing thermodynamic quantities obtained for 2-dim Ising model with exact
solutions.

\subsection{Thermodynamic properties of the $Q=10$ Potts model}

One of advantages of our method is that the density of states does not
depend on temperature; indeed with the density of states, we can calculate
thermodynamic quantities at any temperature. For example, the internal
energy can be calculated by:

\begin{eqnarray}
U(T) = {\frac{{\sum \limits_{E} E g(E) e^{-\beta E} } }{{\sum \limits_{E}
g(E) e^{-\beta E} } }} \equiv \langle E \rangle_T  
\label{eqn:U}
\end{eqnarray}

\begin{figure}[h]
\caption{Thermodynamic quantities calculated from the density of states for
the $Q=10$ Potts model: (a) internal energy, (b) specific heat and the
finite-size scaling function, (c) Gibbs free energy and (d) entropy }
\label{fig:potts_U_1}

(a) \\
\includegraphics{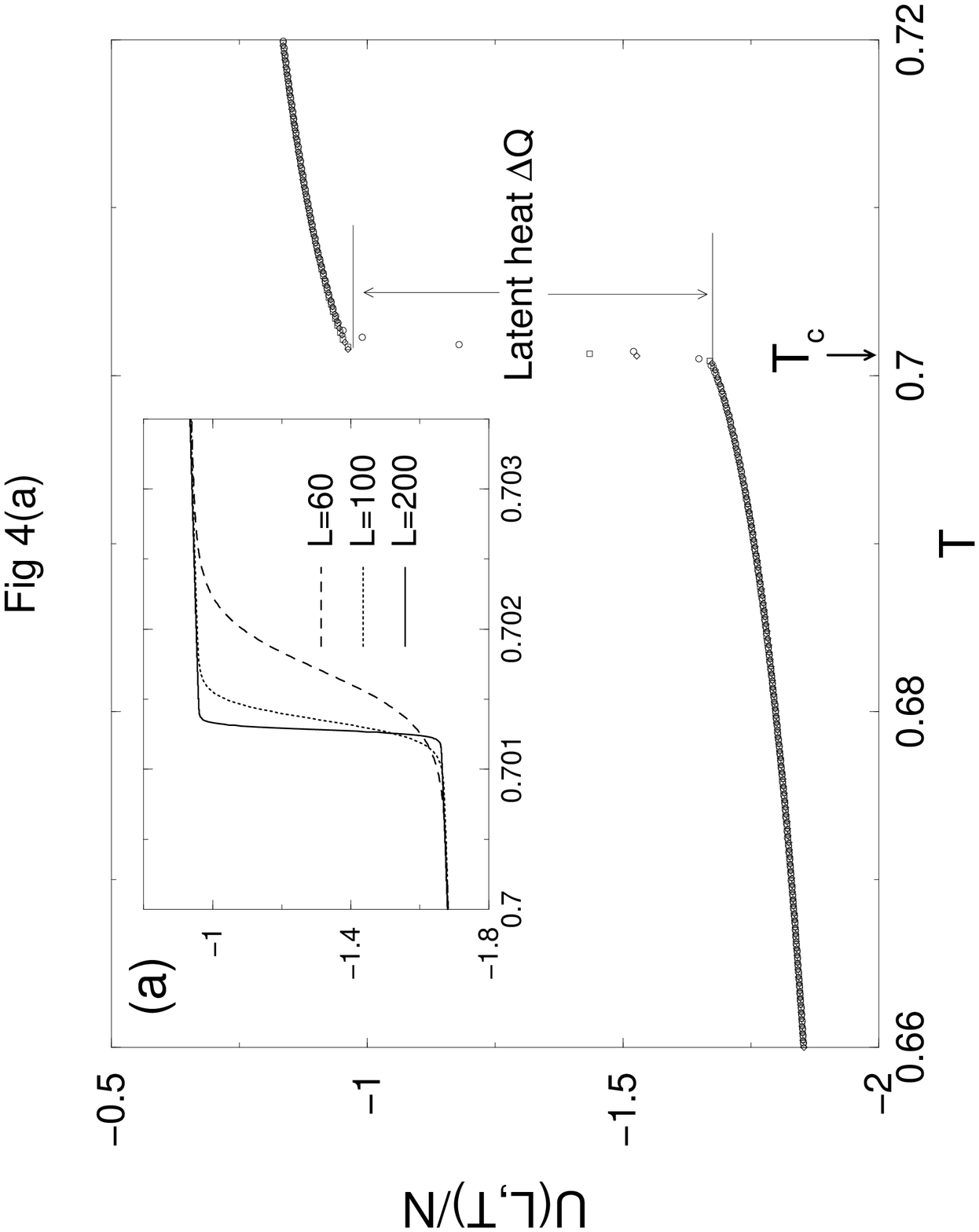}
\vspace{3.0in}

(b) \\
\includegraphics{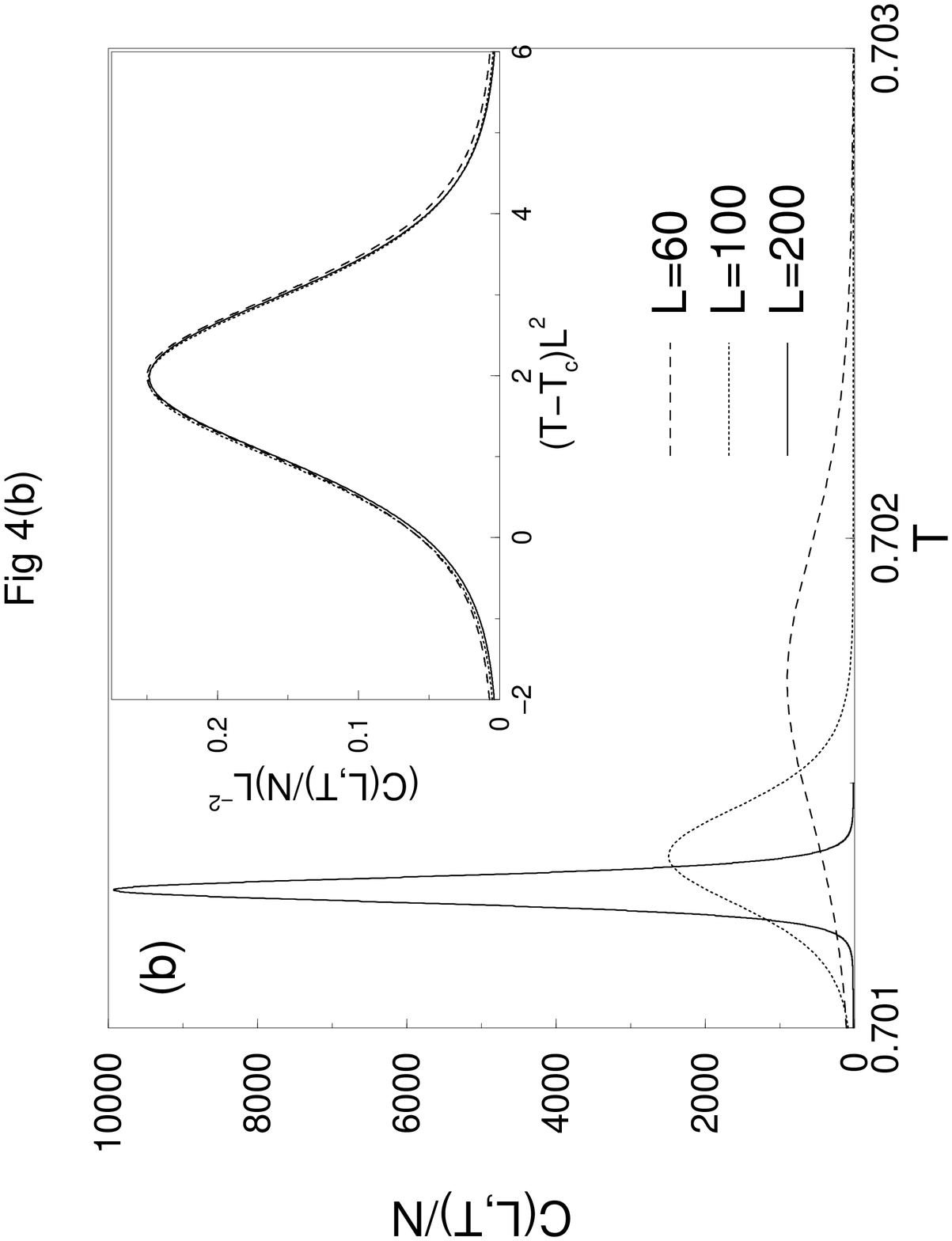}
\vspace{3.0in}

\newpage
(c) \\
\includegraphics{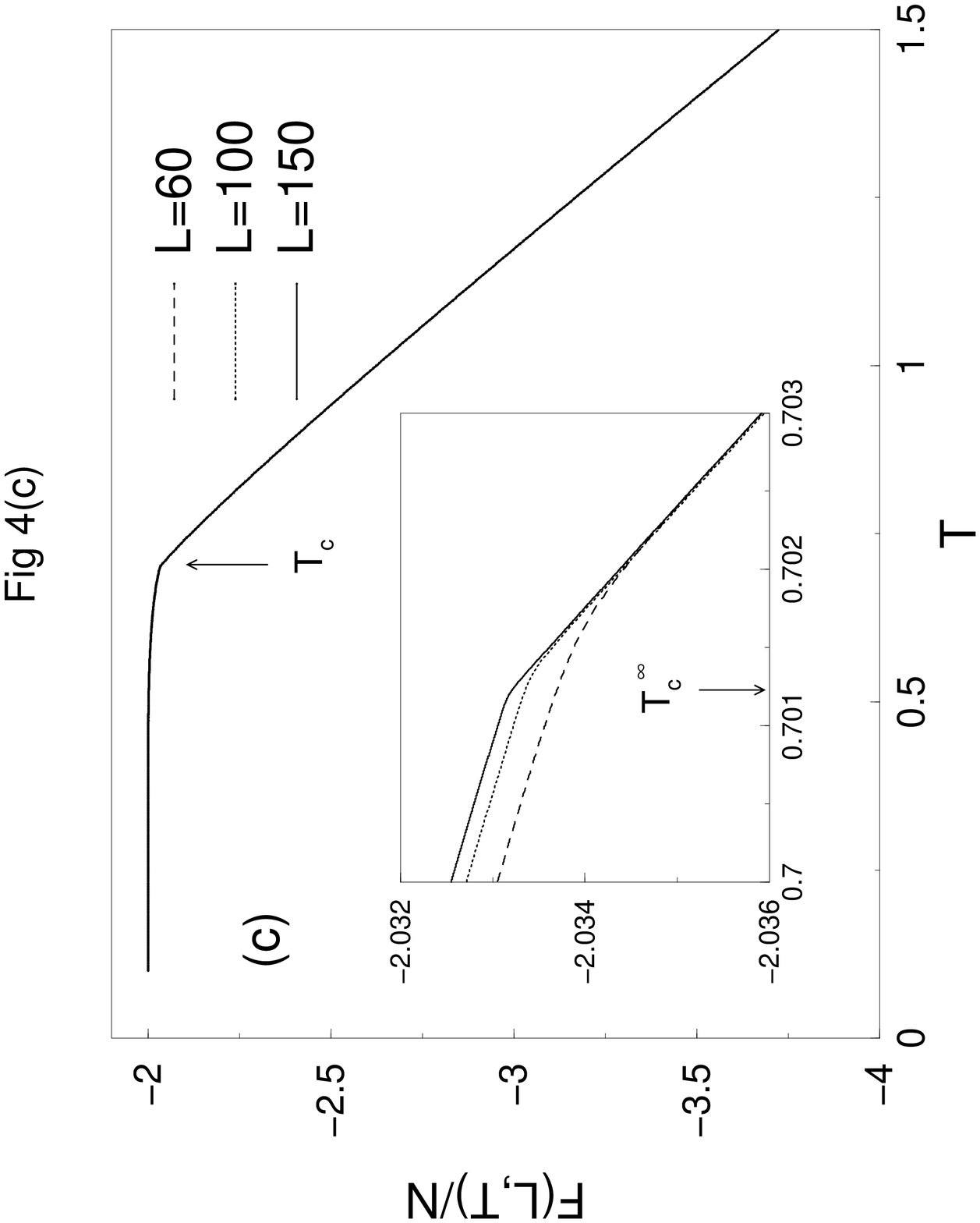}
\vspace{3.0in}

(d) \\
\includegraphics{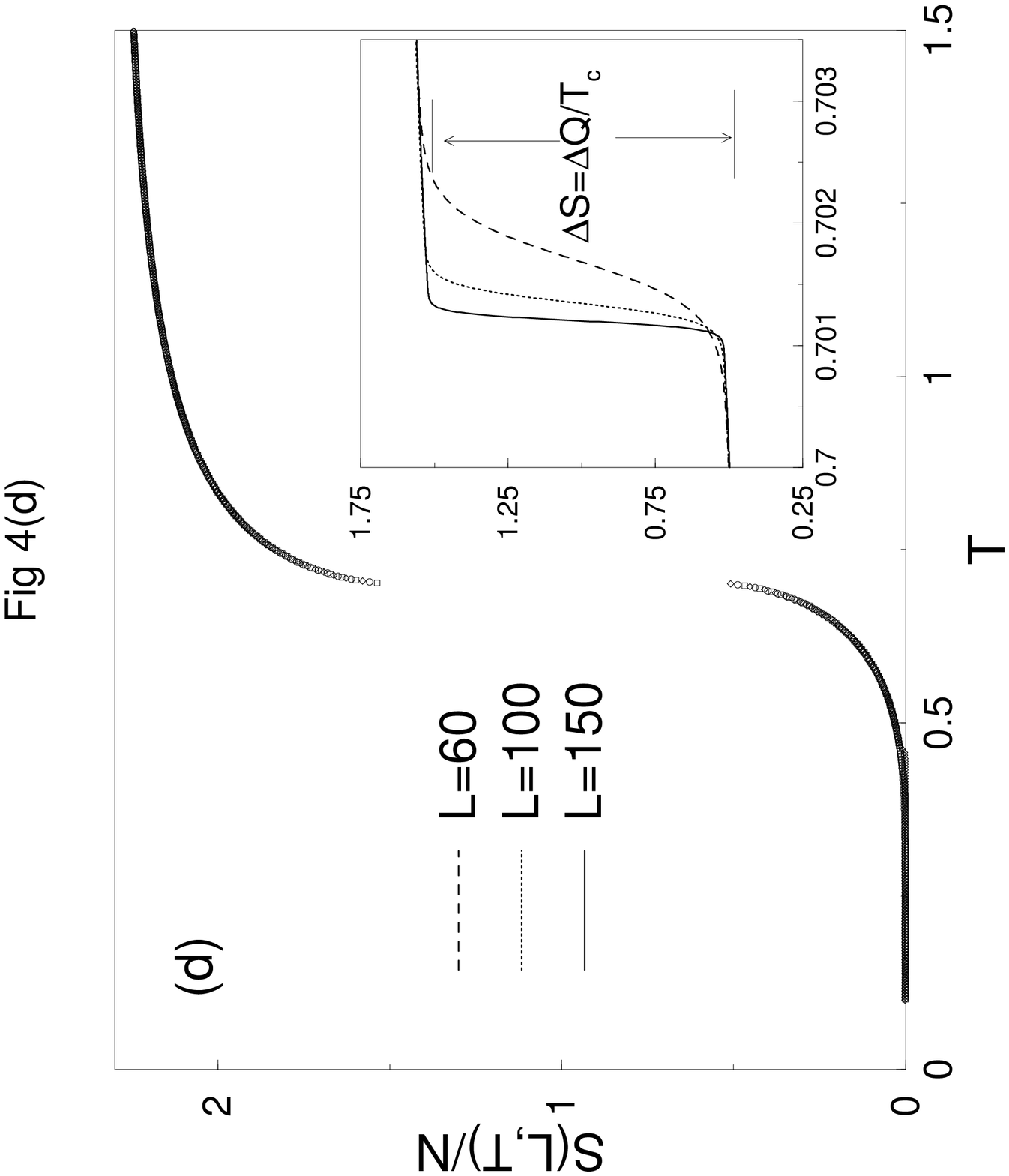}
\vspace{3.0in}

\end{figure}

\noindent To study the behavior of the internal energy near $T_{ext c}$ more
carefully, we calculate the internal energy for $L=60$, 100 and 200
near $T_{\text{c}}$ as presented
in Fig.~\ref{fig:potts_U_1}(a). A very sharp ``jump" in the internal energy
at transition temperature $T_{\text{c}}$ is visible, and the magnitude of
this jump is equal to  the latent heat for the (first-order) phase transition.
Such behavior is related to the double peak distribution of the first-order
phase transition. When $T$ is slightly away from $T_{\text{c}}$, one of the
double peaks increases dramatically in magnitude and the other decreases.

Since we only perform simulations on finite lattices, and use a continuum
function to calculate thermodynamic quantities, all our quantities for
finite-size systems will appear to be continuous if we use a very small
scale. In the inset of Fig.~\ref{fig:potts_U_1}(a) we use the same density
of states again to calculate the internal energy for temperatures very close
to $T_{\text{c}}$. 
On this scale the ``discontinuity'' at the first-order
phase transition disappears and a smooth curve can be seen instead of a
sharp ``jump'' in the main portion of Fig.~\ref{fig:potts_U_1}(a). The
discontinuity in Fig.~\ref{fig:potts_U_1}(a) is simply due to the coarse
scale, but when the system size goes to infinity, the discontinuity will be
real.

From the density of states we can also estimate the specific heat from the
fluctuations in the internal energy: 
\begin{equation}
C(T)={\frac{{\partial U(T)} }{{\partial T }}} = {\frac{{\langle E^2
\rangle_T -{\langle E \rangle_T}^2} }{{T^2} }}  \label{eq:C}
\end{equation}
In Fig.~\ref{fig:potts_U_1}(b), the specific heat so obtained is shown as a
function of temperature. We calculate the specific heat in the vicinity of
the transition temperature $T_{\text{c}}$. The finite-size dependence of the
specific heat is clearly evident. We find that specific heat has a finite
maximum value for a given lattice size $L$ that, according to finite-size
scaling theory for first-order transitions should vary as: 
\begin{equation}
c(L,T)L^{-d} \propto f((T-T_{\text{c}}(\infty))L^{d})
\end{equation}
where $c(L,T)=C(L,T)/N$ is the specific heat per lattice site, $L$ is the
linear lattice size, $d=2$ is the dimension of the lattice. 
$T(L=\infty)=0.70123....$ is the exact solution for the $Q=10$ Potts model~%
\cite{fywu}. In the inset of Fig.~\ref{fig:potts_U_1}(b), our simulational
data for systems with $L=60$, $100$ and $200$ can be well fitted by a single
scaling function, moreover this function is completely consistent with the
one obtained from lattice sizes from $L=18$ to $L=50$ by standard Monte
Carlo~\cite{landau_potts}.

With the density of states, we not only can calculate most thermodynamic
quantities for all temperatures without multiple simulations but can
also access some quantities, such as the Gibbs free energy and entropy,
which are not directly available from conventional Monte Carlo simulations.
The free energy is calculated using 
\begin{eqnarray}
Z &=& \sum\limits_{\{\text{configurations}\}}e^{-\beta E}
=\sum\limits_{E}g(E)e^{-\beta E}  \nonumber \\
F &=& -kT \log(Z)
\end{eqnarray}

Our results for the Gibbs free energy per lattice site is shown in 
Fig.~\ref{fig:potts_U_1}(c) as a function of temperature. Since the transition is
first-order the free energy appears to have a ``discontinuity'' in the first
derivative at $T_{\text{c}}$. This is typical behavior for a first-order
phase transition, and even with the fine scale used in the inset of 
Fig.~\ref{fig:potts_U_1}(c), this property is still apparent even though the system
is finite. The transition temperature $T_{\text{c}}$ is determined by the
point where the first derivative appears to be discontinuous. With a coarse
temperature scale we can not distinguish the finite-size behavior of our
model; however, we can see a very clear size dependence when we view the
free energy on a very fine scale as in the inset of Fig.~\ref{fig:potts_U_1}(c).

The entropy is another very important thermodynamic quantity that cannot be
calculated directly in conventional Monte Carlo simulations. It can be
estimated by integrating over other thermodynamic quantities, such as
specific heat, but the result is not always reliable since the specific heat
itself is not easy to determine accurately, particularly considering the
``divergence" at the first-order transition. With an accurate density of
states estimated by our method, we already know the Gibbs free energy and
internal energy for the system, so the entropy can be calculated easily:

\begin{equation}
S(T) = {\frac{{U(T)-F(T)} }{T}}  \label{eqn:S}
\end{equation}

It is very clear that the entropy is very small at low temperature and at 
$T=0$ is given by the density of states for the ground state. We show the
entropy as a function of temperature in a wide region in Fig. 
\ref{fig:potts_U_1}.(d)

The entropy has a very sharp ``jump'' at $T_{\text{c}}$, just as does the
internal energy and such behavior can be seen very clearly in the inset of
Fig.~\ref{fig:potts_U_1}(d), when we re-calculate the entropy near $T_{\text{c
}}$. The change of the entropy at $T_{\text{c}}$ shown in the figure can be
obtained by the latent heat divided by the transition temperature, and the
latent heat can be obtained by the jump in internal energy at $T_{\text{c}}$
in Fig.~\ref{fig:potts_U_1}(a).

With the histogram method proposed by Ferrenberg and 
Swendsen~\cite{ferrenberg}, it is possible to use simulational data at specific
temperatures to obtain complete thermodynamic information near, or between,
those temperatures. Unfortunately it is usually quite hard to get accurate
information in the region far away from the simulated temperature due to
difficulties in obtaining good statistics, especially for large systems
where the canonical distributions are very narrow. With the algorithm
proposed in this paper, the histogram is ``flat" for the random walk and we
always have essentially the same statistics for all energy levels. Since the
output of our simulation is the density of states, which does not depend on
the temperature at all, we can then calculate most thermodynamic quantities
at any temperature without repeating the simulation. We also believe the
algorithm is especially useful for obtaining thermodynamic information at
low temperature or at the transition temperature for the systems where the
conventional Monte Carlo algorithm is not so efficient.

\subsection{The tunneling time for the $Q=10$ Potts model at $T_{\text c}$}

\begin{figure}[h]
\caption{ Tunneling times $\protect\tau$ for the $Q=10$ Potts model: (a)
comparison of $\protect\tau$ for our random walk algorithm in energy space
and for an ideal random walk in real space; (b) comparison of the tunneling
times for our algorithm, for the multicanonical ensemble method and for the
heat bath algorithm. The solid lines show the ideal case with $\protect\tau%
(N_E) \sim N_E$. }
\label{fig:potts_tt_f}

(a) \\
\includegraphics{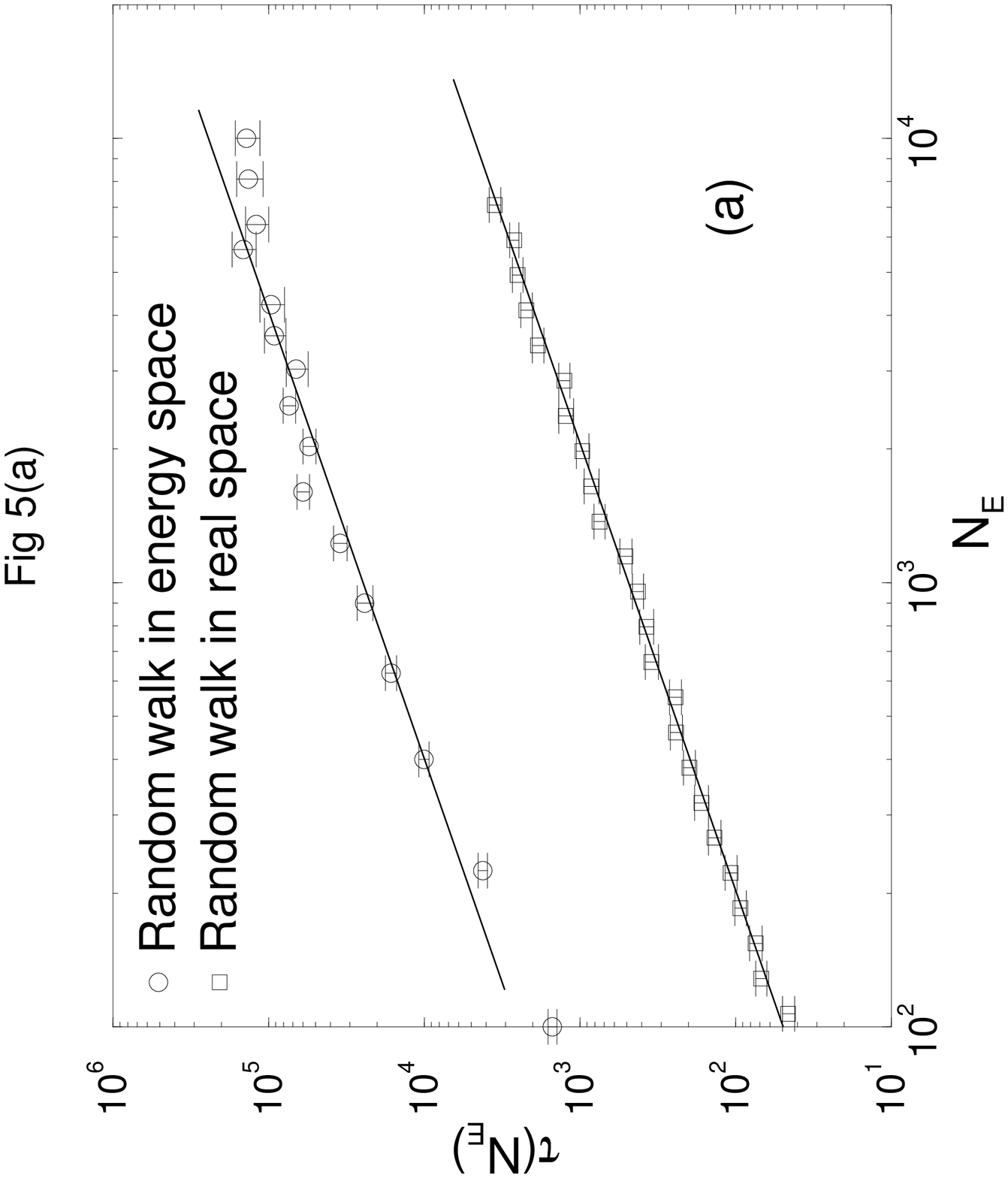}
\vspace{3.0in}

(b) \\
\includegraphics{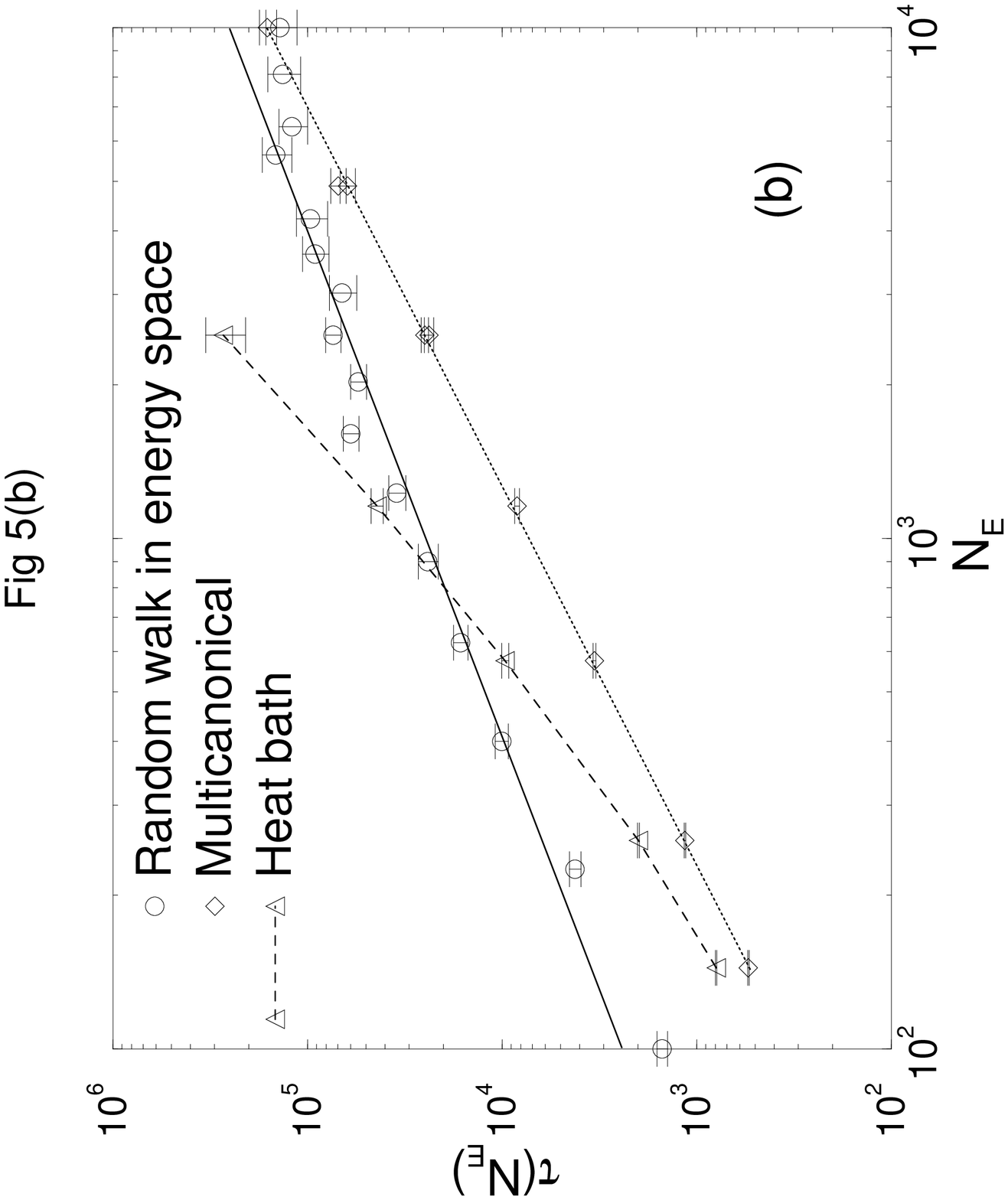}
\vspace{3.0in}

\end{figure}

To study the efficiency of our algorithm, we measure the tunneling time $\tau
$, defined as the average number of sweeps needed to travel from one peak to
the other and return to the starting peak in energy space. Since the
histogram that our random walk produces is flat in energy space, we expect
the tunneling time will be the same as for the ideal case of a simple random
walk in real space, i.e. $\tau(N_E) \sim N_E$, where $N_E$ is the total
number of energy levels. To compare our simulational results to those for
the ideal case, we also perform a random walk in real space. We always use a
fixed $g(x_i)=1$ in one-dimensional real space, where $x_i$ is a discrete
coordinate of position that can be chosen simply as 1, 2, 3 ,4 ...$N_E$. The
random walk is a local random walk with transition probability $p(x_i
\rightarrow x_j)=1/2$, where $x_j=x_{i\pm1}$. We use the same definition of
the tunneling time to measure the behavior of this quantity. The tunneling
time for the ideal case satisfies the simple power law as $\tau(N_E) \sim {
N_E}^\alpha$ and the exponent $\alpha$ is equal to 1. ($\tau$ is defined
using the unit of sweep of $N_E$ sites.) Our simulational data for random
walks in energy space yield a tunneling time that is well described by the
power law $\tau \sim N^{\alpha}$ as shown in Fig.~\ref{fig:potts_tt_f}(a).
The solid lines in the graph have the simple power law as $\tau(L) \sim N_E$
, and we see that our simulation result is very close to the ideal case.
Since our method needs an extra effort to update the density of states to
produce a flat histogram during the random walk in energy space, the
tunneling time is much longer than the real space case. Also because the
tunneling time depends on the accuracy of the density of states, which is
constantly modified during the random walk in energy space, it is not a well
defined quantity in our algorithm. The tunneling time shown in 
Fig.~\ref{fig:potts_tt_f}(a) is the overall tunneling time which includes all
iterations with the modification factors from $f_0=e^1\simeq 2.71828...$ to
the final modification factor $f_{\text{final}}=\exp(10^{-8}) \simeq
1.00000001$.

We should point out that the two processes are not exactly the same, 
since the random walk in real space uses  
the exact density of states($g(x_i)=1$). 
However the random walk in energy space  
requires knowledge of the density of states,
which is {\it a priori} unknown. The algorithm we propose in this paper is a
random walk with the density of states which is modified at each step during
the walk in energy space. At the end of our random walk, the modification
factor approaches 1, and the estimated density of states approaches the true
value. The two processes are then almost identical.

Conventional Monte Carlo algorithms (such as the heat-bath algorithm) have
an exponentially fast growing tunneling time. According to Berg's study in
reference~\cite{berg_1}, the tunneling time obeys the exponential law 
$\tau(L)= 1.46L^{2.15}e^{0.080L}$. The Multicanonical simulational method has
reduced the tunneling time from an exponential law to a power law as 
$\tau(N_E) \sim {N_E}^{\alpha}$. However, the exponent $\alpha$ is as large  as 
$\simeq 1.33$~\cite{berg_0}, which is far away from the ideal case $\alpha=1$.
Only very recently, Janke and Kappler introduced the multibondic cluster
algorithm, the exponent $\alpha$ is reduced to as small as $1.05$ for 2-dim
ten state Potts model~\cite{janke_kappler}. In Fig.~\ref{fig:potts_tt_f}(b),
we show our result with the same quantities obtained with multicanonical
method and heat bath algorithm in reference~\cite{berg_0}. We should point
out that just like multicanonical simulational method, our algorithm has a
power increasing tunneling time with a smaller exponent $\alpha$. 
For small systems, our algorithm offers less  advantage  
because of  the  effort needed to modify the density of states during 
the  random walk. Very recently, Neuhaus has  generalized  this algorithm 
to estimate the canonical distribution for  $T<T_{\text{c}}$,
in magnetization space
for the Ising model~\cite{neuhaus}.  
He found  that for small systems  
the exponent for CPU time versus L
for our algorithm and 
Multicanonical  Ensemble  simulations  are almost  identical.  
However we see that
our algorithm is very efficient for large systems, 
especially  for $L\geq100$.
Our results in Fig.~\ref{fig:potts_tt_f}(b) are  only for 
single range random walks,  
and  multiple range random walks have been 
proven more efficient for  larger systems.

\section{Application to a second order phase transition}

The algorithm we proposed in this paper is very efficient for the study of
any order phase transitions. Since our method is independent of temperature,
it reduces the critical slowing down at the second-order phase transition 
$T_{\text{c}}$ and slow dynamics at low temperature. We estimate the density
of states very accurately with a flat histogram, the algorithm will be a
very efficient for general simulational problems by avoiding the need for
multiple simulations at multiple temperatures.

To check the accuracy and convergence of our method, we apply it to the 2D
Ising model with nearest neighbor interactions on a $L \times L$ square
lattice. This model provides an ideal benchmark for new 
algorithms~\cite{ferrenberg,wang_1} and is also an ideal laboratory for 
testing theory~\cite{beale,landau_ising}. 
This model can be solved exactly, therefore we can
compare our simulational results with exact solutions.

\begin{figure}[h]
\caption{Density of states ($\log(g(E))$) of the 2D Ising model for (a) $%
L=50 $ (single random walks) and (b) $L=256$ (multiple random walks). The
relative errors of the simulational densities of states are shown in the
inset of (a). The overall histogram of the random walk is shown in the inset
of (b). }
\label{fig:ising_density_1}

(a) \\ 
\includegraphics{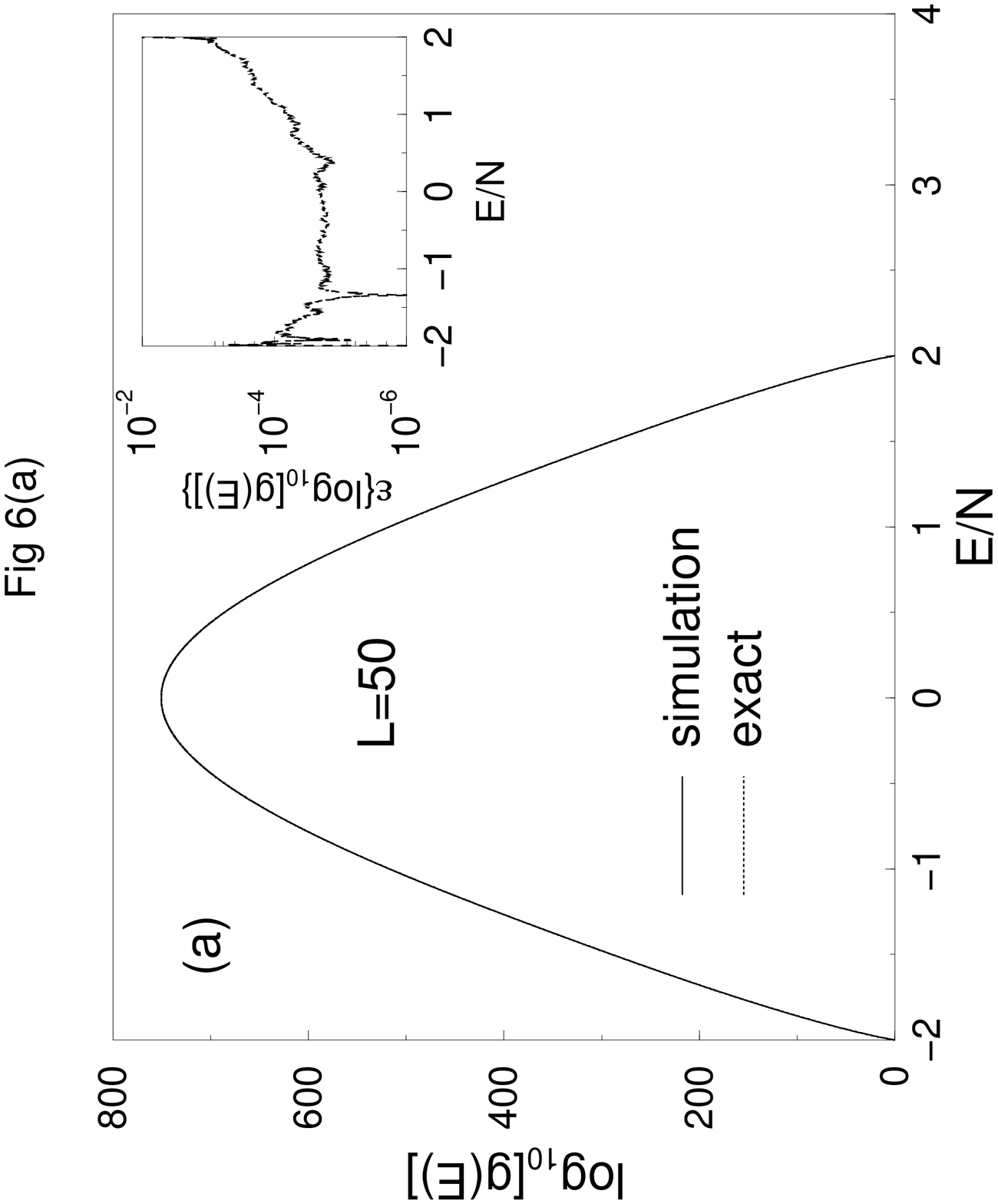}
\vspace{3.0in}

(b) \\
\includegraphics{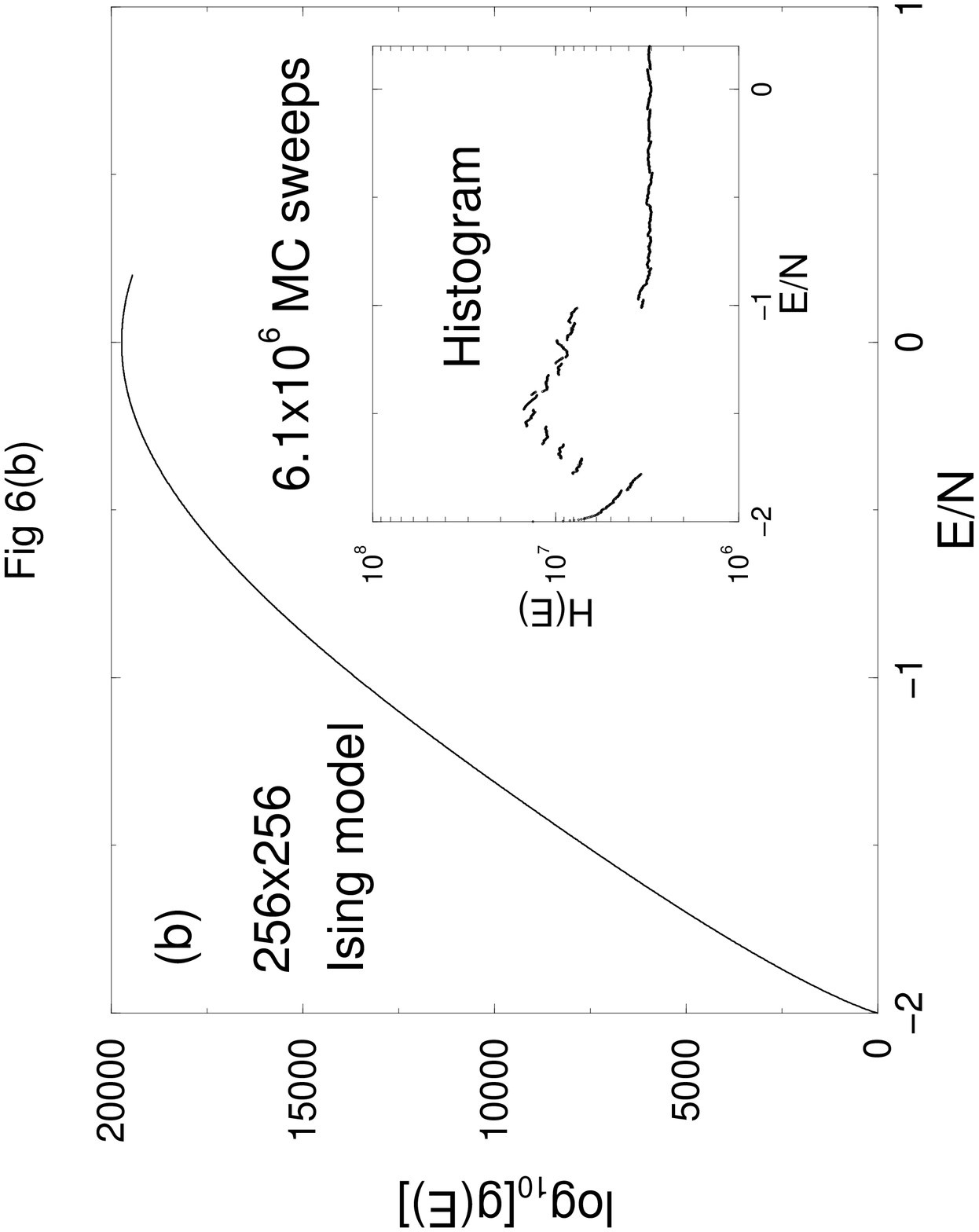}
\vspace{3.0in}

\end{figure}

With the exact solution for the partition function on finite-size 
systems~\cite{ferdinand_fisher}, 
and expansion of the expression by Mathematica, the
density of states for the 2D Ising model on a square lattice can be obtained
exactly~\cite{beale}. Beale~\cite{beale} obtained the exact density of
states up to $L=32$; and because of the memory and speed limitation of
present computers, we could only extend the calculation up to $L=50$ with
the Mathematica program provided by Beale. With the algorithm proposed in
this paper, the density of states for the Ising model is estimated for $L=50$. 
The final modification factor for our random walk was $1.000000001$ for 
$L=50$. In Fig.~\ref{fig:ising_density_1}(a) the simulational densities of
states are compared with exact results. With the logarithmic scale used in
the figure simulational data and exact solution overlap perfectly with each
other for $L=50$. In the inset of the figure, we show the relative error 
$\varepsilon$, which is generally defined by the ratio between error of the
simulational data and exact values for any quantity $X$: 
\begin{equation}
\varepsilon(X) \equiv {\frac {|X_{\text{sim}} -X_{\text{exact}}|} {X_{\text{exact}} } }
\end{equation}
$\varepsilon(\log(g))$ for most of the region is smaller than 0.1\%. Such
errors for low energy levels are directly related to the errors for the
thermodynamic quantities calculated from the density of states. The average
relative error is 0.019\% for $L=50$. It is possible to estimate the density
of states for small systems with the broad histogram 
method~\cite{oliveira_0,oliveira_1,oliveira_2}. Recent broad histogram simulational 
data~\cite{Lima} for the 2D Ising model on a $32\times 32$ lattice with $10^{6}$
MC sweeps yielded an average deviation of the microcanonical entropy of  
about 0.08 \% from the exact solution~\cite{beale}. With our algorithm we
obtain an average error as small as 0.035 \% on the $32\times 32$ lattice
with $7\times 10^{5}$ MC sweeps.
Procedures for allowing  $f \rightarrow 1$ have been  examined  by 
H\"uller~\cite{hueller} who used data from two densities of states
for two different values of $f$ to extrapolate  to $f=1$. 
However, his data for a small Ising system  yield larger
errors than  our direct approach. 
The applicability of his method to  large systems 
also needs a more detailed  study.

The absolute density of states in Fig.~\ref{fig:ising_density_1}(a) is
obtained by the condition that the number of ground states is 2 for the 2D Ising
model (all up or down). This condition guarantees the accuracy of the density
of states at low energy levels which are very important in the calculation
of thermodynamic quantities at low temperature. With this condition, when 
$T=0$, we can get exact solutions for internal energy, entropy and free
energy when we calculate such quantities from the density of states. If we
apply the condition that the total number of states is $2^N$ for the
ferromagnetic Ising model, we can not guarantee the accuracy of the energy
levels at or near ground states because the rescaled factor is dominated by
the maximum density of states.

In Fig.~\ref{fig:ising_density_1}(b), we show our estimation of the density
of states of Ising model on $256\times 256 $ lattice. Since the density of
states for $E>0$ has almost no contribution to the canonical average at finite
positive temperature, we only estimate the density of states in the region 
$E/N\in[-2,0.2]$ out of the whole energy $[-2,2]$. To speed up our
calculation, we divide the desired energy region [-2, 0.2] into 15 energy
segments, and estimate the density of states for each segment with
independent random walks. The modification factor changes from 
$f_0=e^1\simeq2.71828...$ to $f_{\text{final}}=\exp(10^{-7})\simeq
1.0000001...$. The resultant density of states can be joined from adjacent
energy segments. To reduce the boundary effects of the random walk on each
segments, we keep about several hundred overlapping energy levels for random
walks on two adjacent energy segments. The histograms of random walks are
shown in the inset of this figure. We only require a flat histogram for 
each energy segment. To reduce the error of the density of states relevant
to the accuracy of the thermodynamic quantities near $T_{\text c}$ 
we optimize the
parameter and perform additional multiple random walks for the energy range 
$E/N \in [-1.8, -1]$ with same number of processors. For this we use the
density of states obtained from the first simulations as starting points and
continue the random walk with modification factors changing from 
$\exp(10^{-6})\simeq 1.000001$ to $\exp(10^{-9})\simeq 1.000000001$. The
total computational effort is about $9.2 \times 10^6$ visits on each energy
levels. Note that the total number of possible energy levels is $N-1$ and we
perform random walks only on [-2, 0.2] out of [-2, 2]. The real simulational
effort is about $6.1 \times 10 ^6$ MC sweeps for the Ising model with $L=256$.
With the program we implemented, it took about 240 CPU hours on a single
IBM SP Power3 processor.

\begin{figure}[h]
\caption{Boundary effects for random walks in different energy ranges for
the $2D$ Ising model. (a) errors in the density of states; (b) specific
heats calculated from the density of states; and (c) the specific heat if
the two highest energy entries in the density of states are deleted. }
\label{fig:ising_density_error_range}

(a) \\
\includegraphics{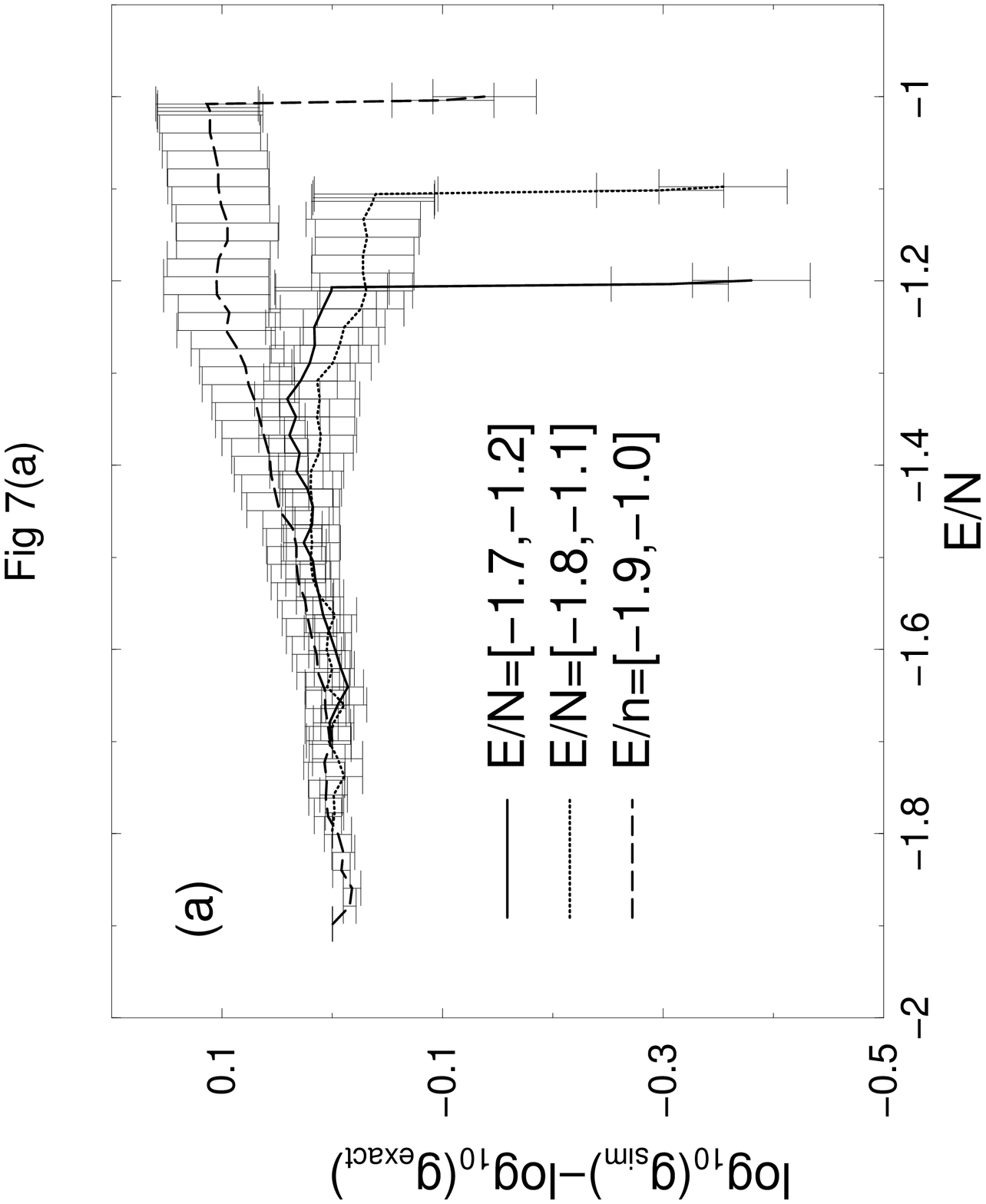}
\vspace{3.0in}

(b) \\
\includegraphics{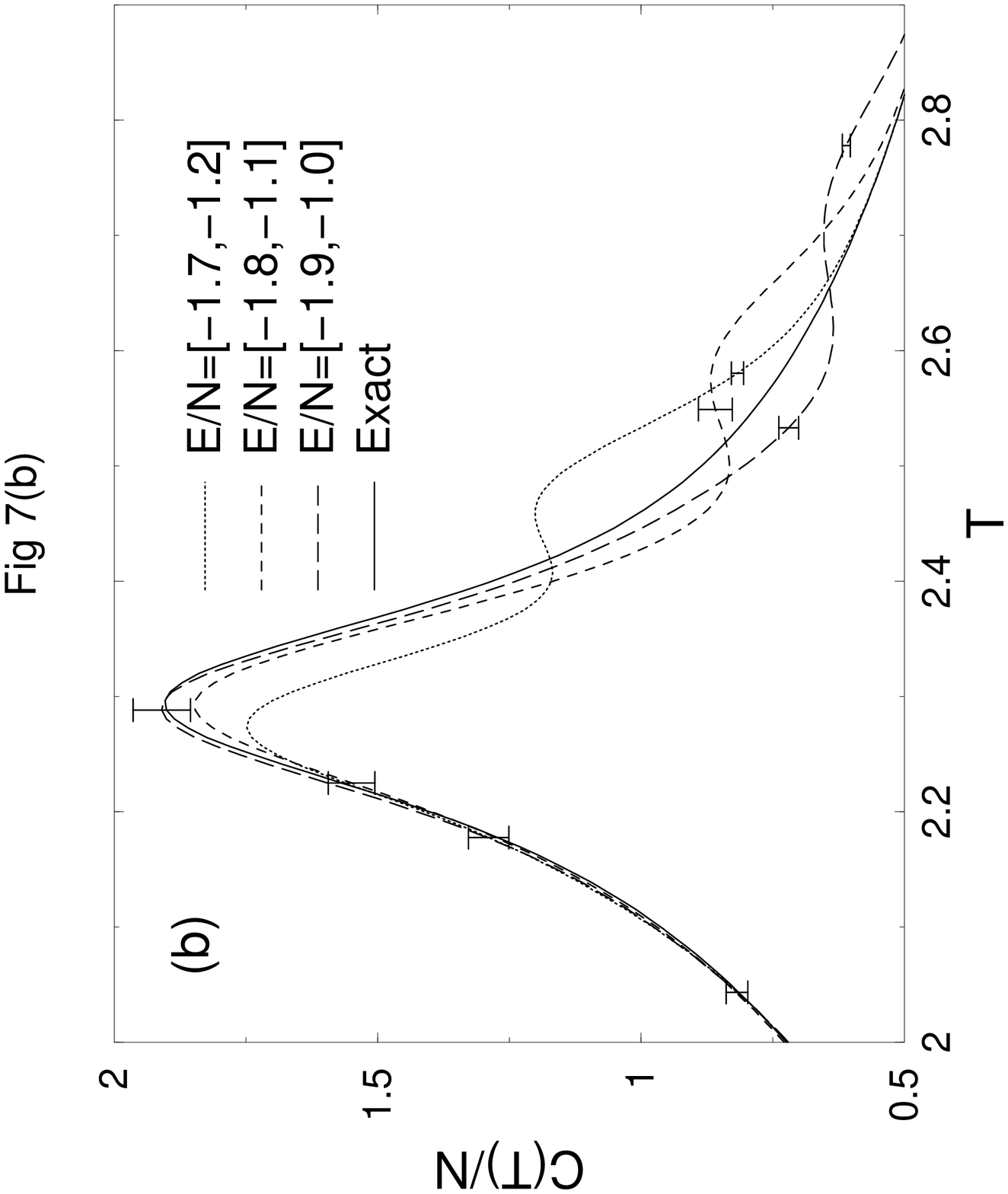}
\vspace{3.0in}

\newpage

(c) 
 
\includegraphics{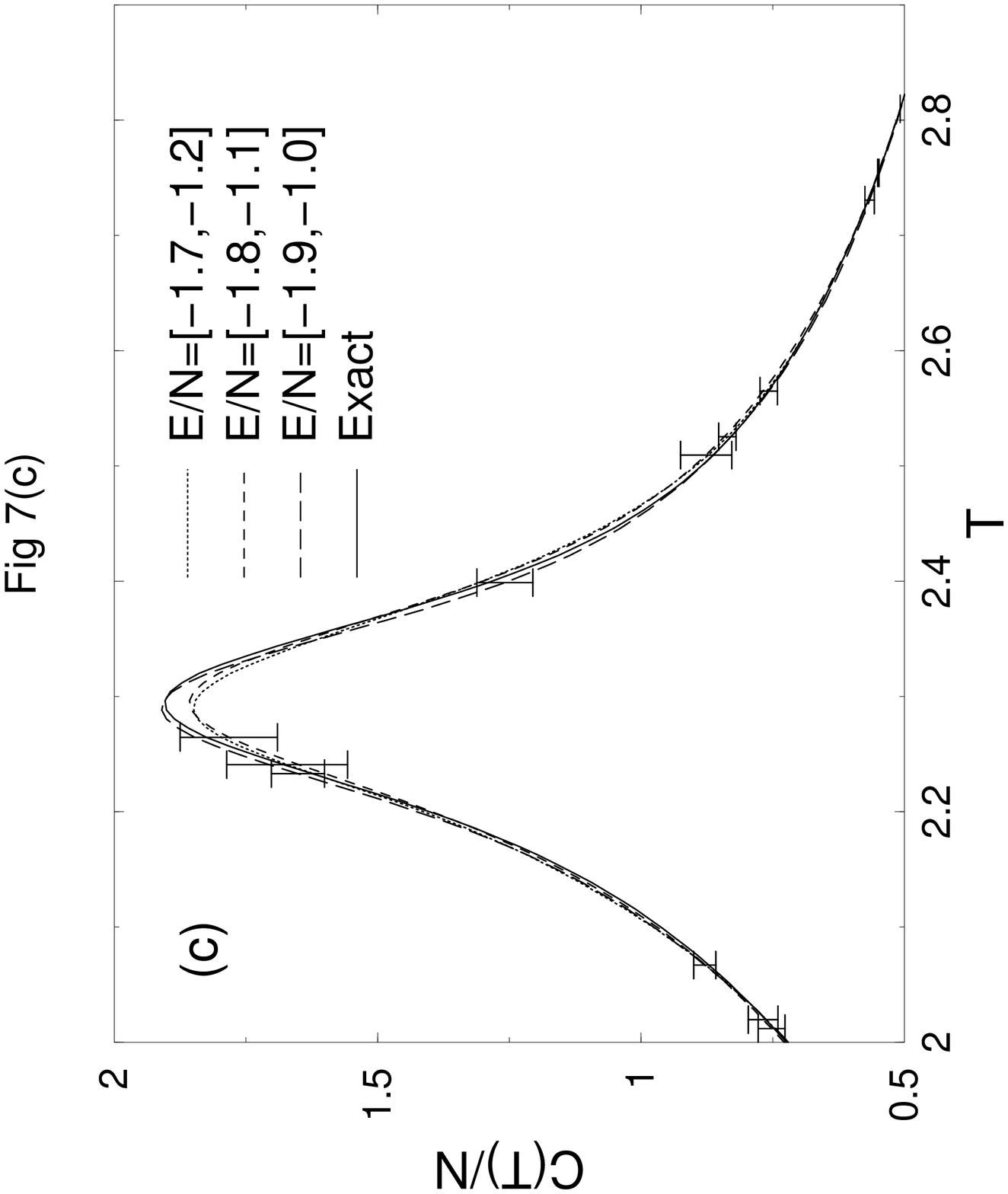}
\vspace{3.0in}

\end{figure}

For $L=256$, we perform multiple random walks on different energy ranges,
and one problem arises, that is the error of the density of states due to
the random walk in a restricted energy range. Since the exact density of
states for large systems is not available, we use $L=32$ to study such
effects. We perform three independent random walks in the ranges $E/N=[-1.7,
-1.2]$, $E/N=[-1.8, -1.1]$ and $E/N=[-1.9,-1.0]$ to calculate the densities
of states on these ranges. In Fig.~\ref{fig:ising_density_error_range}(a),
we show the errors of our simulation results from the exact values. We make
our simulational densities of states match up with the exact results at the
left edges. It is very clear that the width of the energy range of the
random walks is almost not relevant to the errors of the density of states.
The reason is that the random walks only require the local histogram
to be  flat as
we discussed in the previous section. But the errors for the last two densities
on the right edges are significantly larger than the others. The problem is
the boundary effect because in this simulation we treat the densities of
states at edges as same as those away from edges. That is not correct.
Another reason is  due to the flat histogram and high density at right
side edges. Our simulational effort for the densities at the right edges is
not enough compared to those at the left side. This is also the reason why
we have not seen big errors at left edges.

To study the influence of the errors of the densities of states on the
thermodynamic quantities calculated from them, in the energy range that we
perform random walks, we replace the exact density of states with the
simulational density of states. In Fig.~\ref{fig:ising_density_error_range}%
(b), the specific heat calculated from such density of states is showed as a
function of temperature. We also show the exact value with the simulational
data, the difference is obvious.

To reduce the boundary effect, we delete the last two density entries, and
insert them into the exact density of states again, then the results in 
Fig.~\ref{fig:ising_density_error_range}(c) are much better. 

With our test in the three different ranges of energy, it is quite safe to
conclude that boundary effect will not be present in our multiply random walks
if we have a couple of energy levels overlap for adjacent energy ranges. In
our real simulations for large systems, we have hundreds of overlaping energy
levels.

Since the exact density of states is only available on small systems
($ L\leq50 $), it is not so interesting to compare the simulational density of
states itself. The most important thing is the accuracy of estimations of
thermodynamic quantities calculated from such density of states on large
systems. With the density of states on large systems, we apply canonical
average formulas to calculate internal energy, specific heat, Gibbs free
energy and entropy. Ferdinand and Fisher~\cite{ferdinand_fisher} obtained
the exact solutions of above quantities for 2D Ising model on finite-size
lattices. Our simulational results on finite-size lattice can be compared
with those exact solutions.

\begin{figure}[h]
\caption{Thermodynamic quantities for the $2D$ Ising model calculated from
the density of states. Relative errors with respect to the exact solutions
by Ferdinand and Fisher are shown in the insets. (a) Internal energy, (b)
specific heat, (c) Gibbs free energy and (d) entropy }
\label{fig:ising_U}

(a)\\ 
\includegraphics{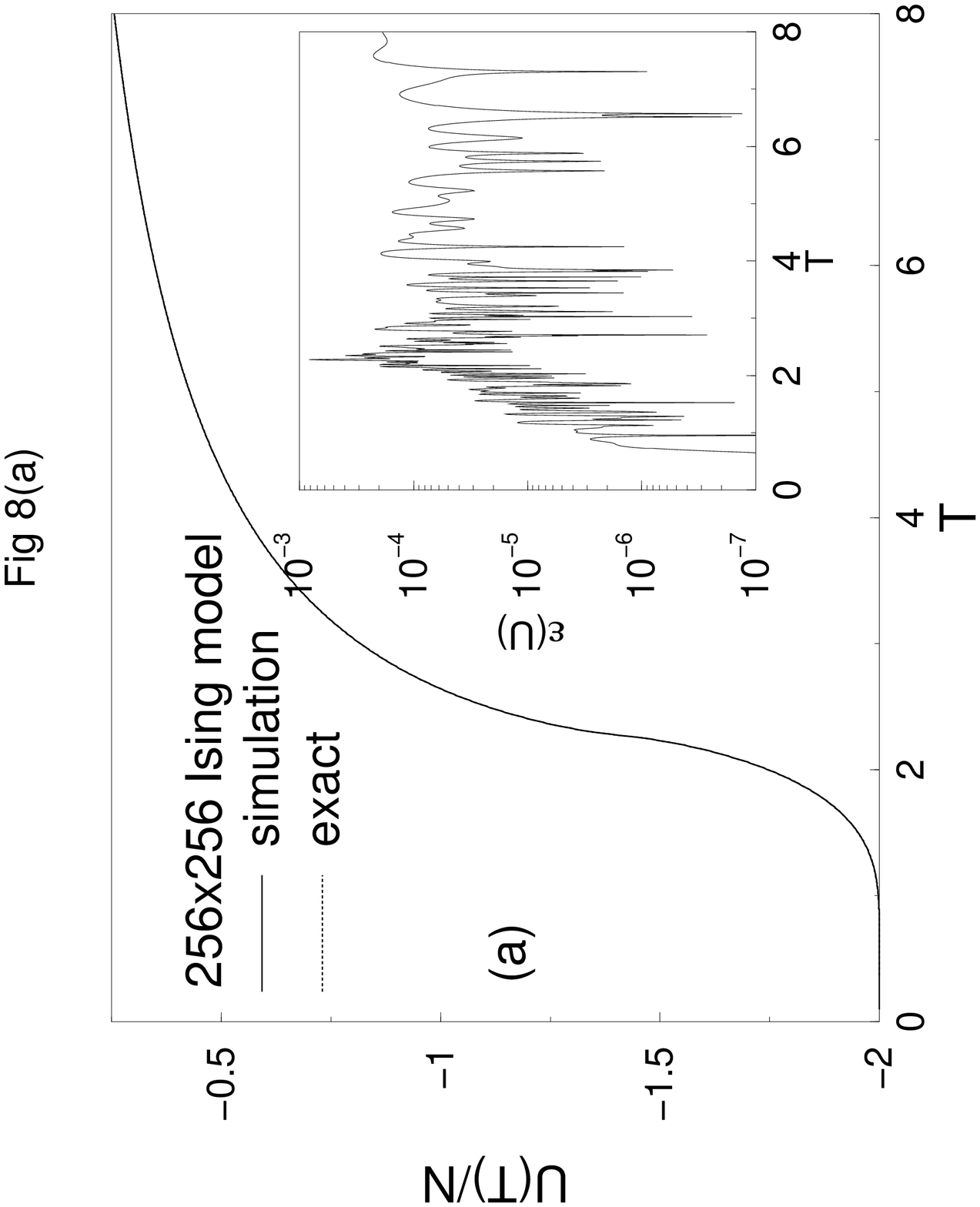}
\vspace{3.0in}

(b)\\  
\includegraphics{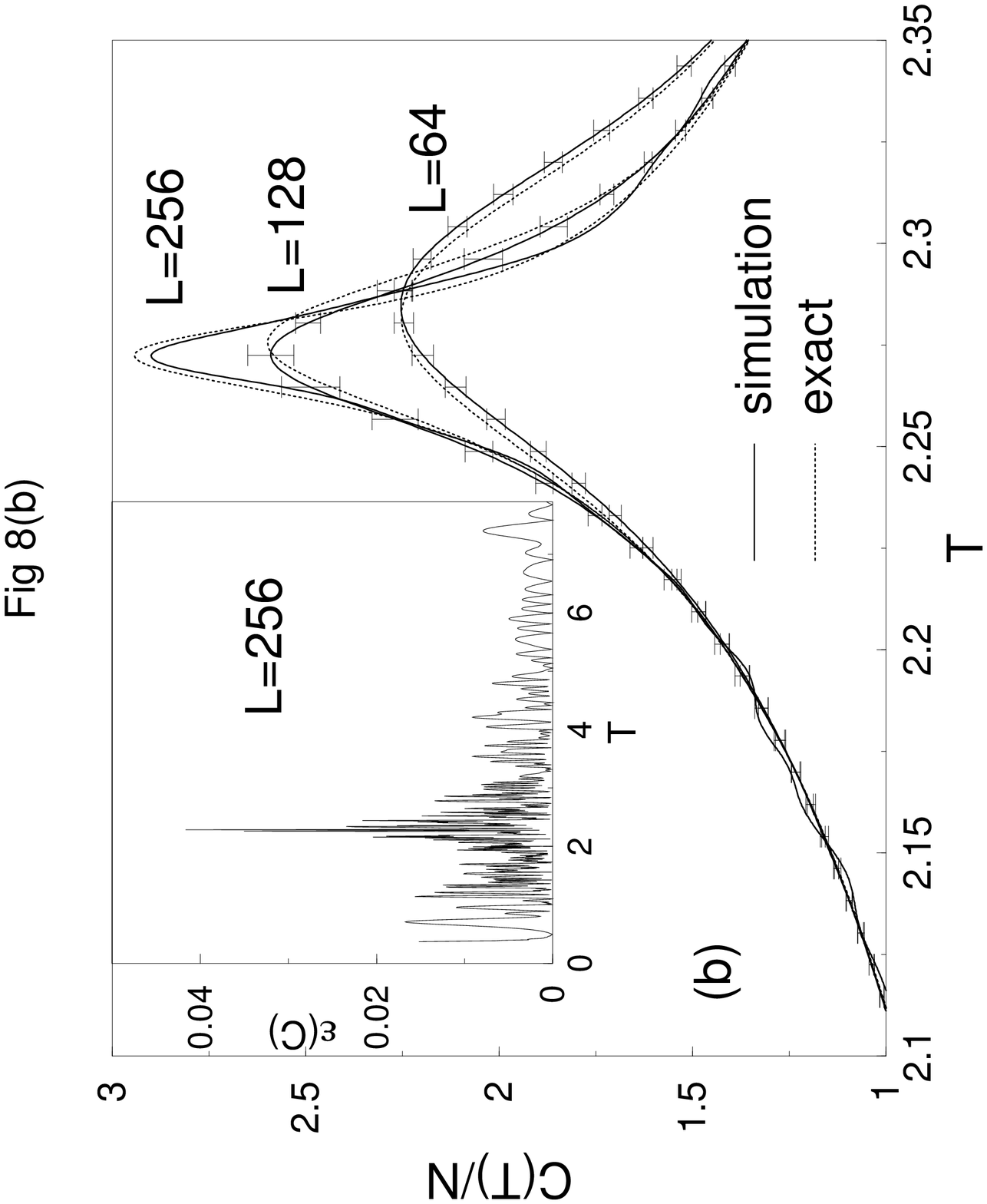}
\vspace{3.0in}

\newpage

(c) \\ 
\includegraphics{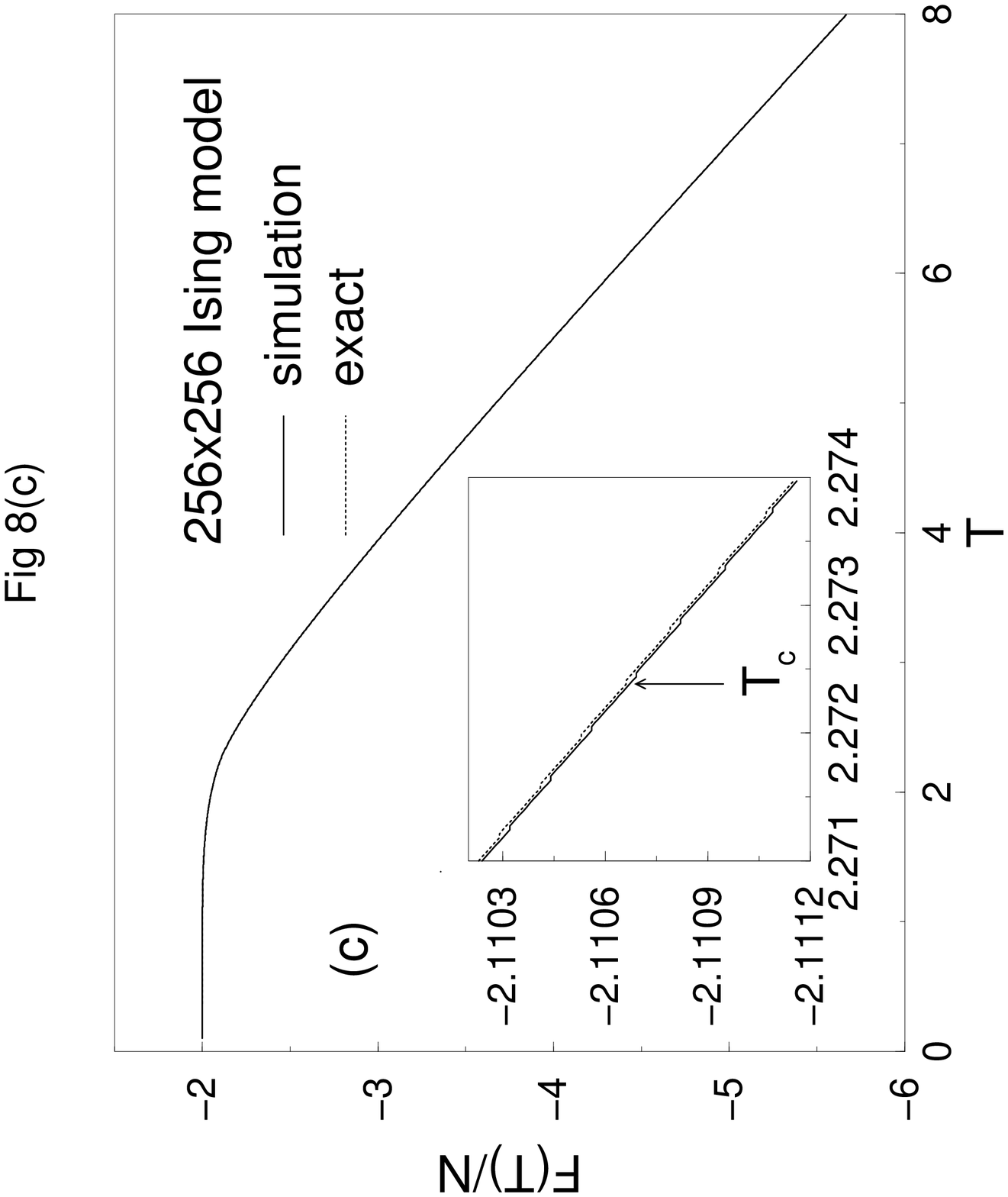}
\vspace{3.0in}

(d) \\  
\includegraphics{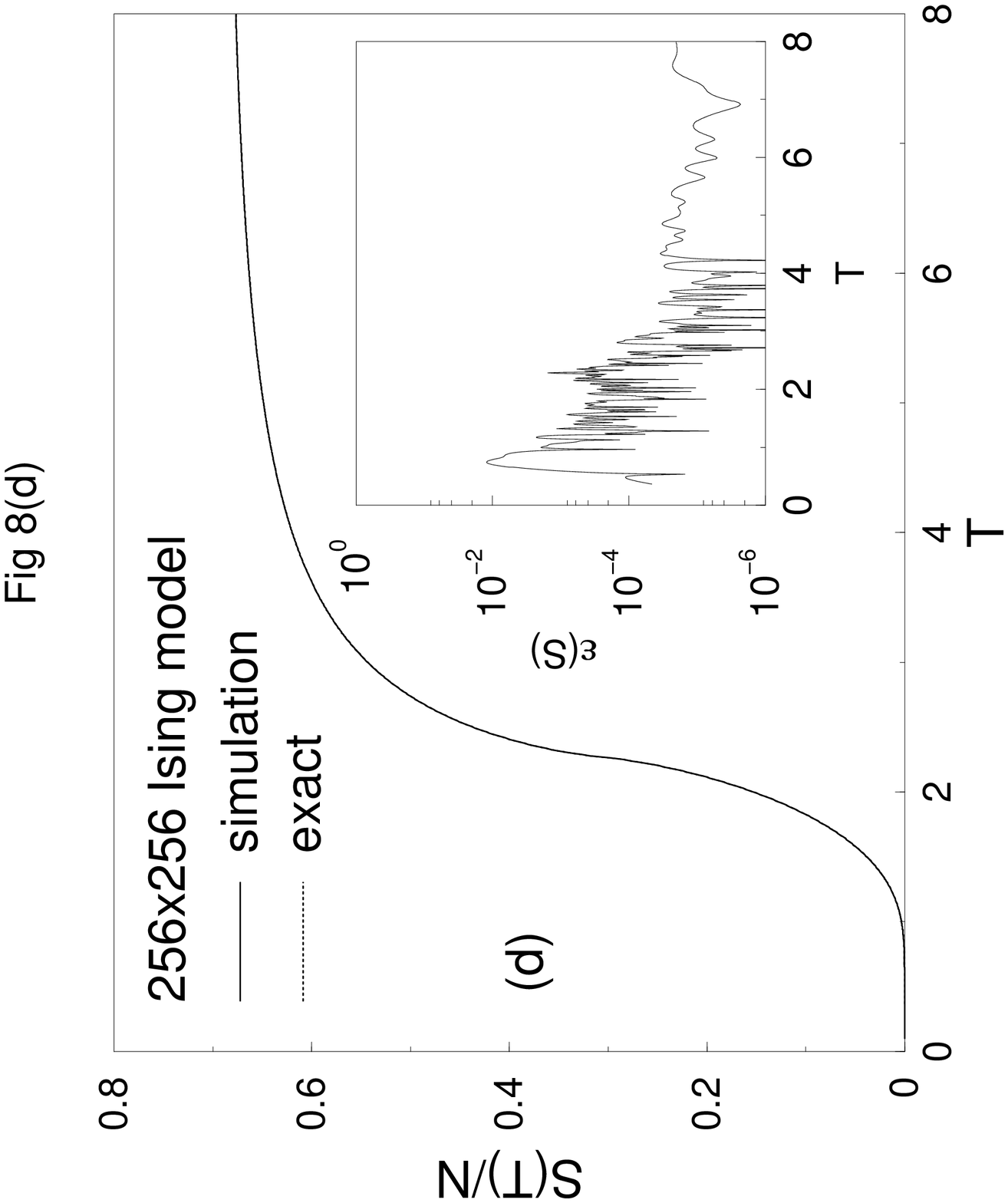}
\vspace{3.0in}

\end{figure}

In Fig.~\ref{fig:ising_U}(a), we show the internal energy as a function of
temperature. The internal energy is estimated from the canonical average
over energy of the system as the equation (\ref{eqn:U}). We also draw the
exact solution in the same graph. The exact and simulational data perfectly
overlap with each other in a wide temperature region from $T=0$ to $T=8$.
Since no difference is visible with the scale used in Fig.~\ref{fig:ising_U}%
(a), a more stringent test of the accuracy is provided by the inset which
shows the relative errors $\varepsilon(U)$.  With the density of states
obtained with our algorithm, the relative error of simulational internal
energy for $L=256$ is smaller than 0.09\% for the temperature region from 
$T=0$ to $8$.
From  eqn. (\ref{eqn:U})  
it is very clear that  the canonical distribution 
serves as  a weighting factor, and since  
the distribution is very narrow,   
$U(T)$ is only determined by a small portion 
of the density of states. 
(For the $L=50$ 2D Ising model at $T_{\text c}$, 
only the density of states for $E/N \in [-1.6, -1.2]$ 
contributes in a  major way to  the calculation.)   
Therefore  the error $\varepsilon(U)$  
is also determined by the 
errors  of the density of states in the same narrow  energy  range. 

With the density of states and equation (\ref{eq:C}), we also can calculate
the specific heat per lattice site as a function of temperature. Both
simulational data and exact results near $T_{\text c}$ are shown in Fig.~\ref%
{fig:ising_U}(b) for $L=64$, 128 and 256 Ising model in the vicinity of $T_{\text c}$. 
Within the resolution
of the figure, we can see the difference between the simulational data and
exact solutions. In the inset of Fig.~\ref{fig:ising_U}(b), we present
relative errors for our simulational data as a function of temperature for $%
L=256$. The errors for the specific heat for the Ising model 
on a $256 \times 256
$ lattice are smaller than $4.5 \%$ in all temperature region $T<8$. 
Very
recently, Wang, Tay and Swendsen~\cite{wang_1} estimated the specific heat
of the same model on a $64\times 64$ lattice by the transition matrix Monte
Carlo re-weighting method~\cite{swendsen}, and for a simulation with $%
2.5\times 10^{7}$ MC sweeps, the maximum error in temperature region $T\in
\lbrack 0,8] $ was about 1\%. When we apply our algorithm to the same model
on the $64\times 64$ lattice, with a final modification factor of $%
1.000000001$ and a total of $2\times 10^{7}$ MC sweeps on single processor,
the errors in the specific heat are reduced below 0.7\% for all temperatures.
The relatively large errors at low temperature reflect the small values for
the specific heat at low temperature. According to the  simulational data
by broad histogram method, the errors in specific heat are very large even
for systems as small as $32 \times 32$~\cite%
{oliveira_0,oliveira_1,oliveira_2}. Only very recently, they have reduced
the error near $T_{\text c}$ to a small value for $L=32$~\cite{oliveira_5}.

In Fig.~\ref{fig:ising_U}(c), the free energy for $L=256$ is plotted as a
function of temperature. For comparison the exact solution~\cite%
{ferdinand_fisher} is shown in the same figure. As expected, simulational
and exact data  overlap perfectly within the resolution of the figure.
Since the system has a second-order phase transition, unlike the $Q=10$
Potts model, the first derivative of the free energy is a continuous
function of temperature. The result very close to $T_{\text c} (L=256)$ is
shown in the inset of Fig.~\ref{fig:ising_U}(c) for $L=256$. The behavior of
the free energy near $T_{\text c}$ is quite different from that of the
first-order phase transition in Fig.~\ref{fig:potts_U_1}(c). 
According to our calculation, the relative errors for all temperatures from $%
T=0$ to $8$ are smaller than 0.0008\%, which means that our simulational data
agree almost perfectly with the exact solution~\cite{wang_landau}. Since we
use the condition that the number of ground states is $g(E=-2N)=2$ to
normalize the density of states, the errors at low temperatures are
extremely small.

The entropy of the 2D Ising model can be calculated with the equation(\ref%
{eqn:S}). In Fig.~\ref{fig:ising_U}(d), the simulational data and exact
results are presented in the same figure. With the scale in the figure, the
difference between our simulational data and exact solutions are not
visible. In the inset of Fig.~\ref{fig:ising_U}(d), the relative errors of
our simulational data are plotted as a function of temperature. For the
Ising model on a $256 \times 256 $ lattice, the relative errors are smaller
than 1.2\% for all temperature range. We also notice that the errors near $%
T=0$ decrease dramatically. The reason is the condition we use to normalize
the density of states. Very recently, with the flat histogram method~\cite%
{wang_3} and the broad histogram method~\cite%
{oliveira_0,oliveira_1,oliveira_2}, the entropy was estimated with $10^{7}$
MC sweeps for the same model on $32\times 32$ lattice; however, the errors
in reference~\cite{wang_0} are even much bigger than our errors for $%
256\times 256$!

\section{Application to 3D $\pm J$ EA model}

Spin glasses~\cite{Binder} are magnetic systems in which the interactions
between the magnetic moments produce frustration because of some structural
disorder. One of the simplest theoretical models for such systems is the
Edwards-Anderson model~\cite{Edwards} (EA model) proposed twenty five years
ago. For such disordered systems, analytical methods can provide only very
limited information, so computer simulations play a particularly important
role. However, because of the rough energy landscape of such disordered
systems, the relaxation times of the conventional Monte Carlo simulations
are very long. The dynamical critical exponent was estimated as large as $%
z\simeq 6$~\cite{Ogielski1,fgwang1,fgwang2}. Normally simulations can be
performed only on rather small systems, and many properties concerning the
spin glasses are still left unclarified~\cite%
{Bhatt1,Kawashima,Palassini_1,Palassini_2,Palassini_3,Marinari_1,Moore,Houdayer}%
.

In this paper, we consider the three-dimensional  $\pm J$ 
Ising spin glass EA  model.
The model is defined by the Hamiltonian 
\begin{equation}
{\cal H} =-\sum_{\langle i,j \rangle}J_{ij}\sigma_i\sigma_j
\end{equation}
where $\sigma$ is an Ising spin  and the coupling   
$J_{ij}$ is quenched to $\pm 1$ randomly.  
The summation runs over the nearest-neighbors $\langle i,j\rangle$  
on a simple cubic lattice.   

One of most important issues for a spin glass model is the low temperature
behavior. Because of the slow dynamics and rough phase space landscape of
this model, it is also one of most difficult problems in simulational
physics. The algorithm proposed here is not only very efficient in
estimating the density of states but also very aggressive in finding the
ground states. From a random walk in energy space, we can estimate the
ground state energy and the density of states very easily. For a spin glass
system, after we finish the random walk, we can obtain the absolute density
of states by the condition that total number of states is $2^N$. The entropy
at zero temperature can be calculated from either $S_0=\ln (g(E_0))$ or $%
\lim\limits_{T \rightarrow 0}{\frac {U-F}{T}}$, where $E_0$ is the energy at ground
states. Both relations will give the same result since $U$ and $F$ are
calculated from the same density of states. Our estimates for $s_0=S_0/N$
and $e_0=E_0/N$ per lattice site, listed in Table 2, agree with the
corresponding estimates made with the multicanonical method. With our
algorithm, we can estimate the density of states up to $L=20$ by a random
walk in energy space for few hours on a 400MHz processor.

If we are only interested in the quantities directly related to the energy,
such as free energy, entropy, internal energy and specific heat, one
dimensional random walk in energy space will allow us to calculate these
quantities with a high accuracy as we did in the 2D Ising model. However for
spin glass systems, one of the most important quantities is the order
parameter which can be defined by~\cite{Edwards} 
\begin{equation}
q^{\text {{\text EA}}}(T) \equiv \lim _{t \rightarrow \infty} \lim _{N
\rightarrow \infty} q(T,t), \quad q(T,t) \equiv \langle
\sum_{i=1}^{N}\sigma_{i}(0)\sigma_{i}(t)/N \rangle.
\end{equation}
Here, $N=L^3$ is the total number of the spins in the system, $L$ is the
linear size of the system, $q(T,t)$ is the auto-correlation function, which
depends on the temperature $T$ and the evolution time $t$, and $q(T,0)=1$.
When $t\rightarrow \infty$, $q(T,t)$ becomes the order parameter of the spin
glass. This parameter takes the following values
\begin{equation}
q^{\text{EA}}(T) \left\{ 
\begin{array}{c}
=1 \ {\text {if}} \ T=0 \\ 
= 0 \ {\text  {if}}\  T \geq T_{\text {g}} \\ 
\neq 0 \ {\text {if}} \ 0<T<T_{\text { g}}%
\end{array}
\right.,
\end{equation}
The value at $T=0$ can be different from 1 in the case where the ground
state is highly degenerate.

In our simulation, there is no temperature introduced during the random
walk. And it is more efficient to perform a random walk in single system
than two replicas. So the order-parameter can be defined 
\begin{equation}
q\equiv \langle \sum_{i=1}^{N}\sigma_{i}^{0}\sigma_{i}/N \rangle.
\end{equation}
where $\{\sigma_{i}^{0}\}$ is one of spin configurations at ground states and $%
\{\sigma_{i}\}$ is any configuration during the random walk. The behavior of $q$
we defined above is basically the same as the order-parameter defined by the
Edwards and Anderson~\cite{Edwards}. It is not exact same order-parameter
defined by Edwards and Anderson, but was used in the early numerical
simulations by Binder {\it et al.}\cite{binder_q_1,binder_q_2}.

After first generating a bond configuration we perform a one-dimensional
random walk in energy space to find a spin configuration $\{\sigma_i^{0}\}$ for
the ground states. Since the order-parameter is not directly related to the
energy, to get a good estimate of this quantity we have to perform a
two-dimensional random walk to obtain the density of states $G(E,q)$ with a
flat histogram in $E$-$q$ space. This also allows us to overcome the
barriers in parameter space (or configuration space) for such a complex
system. The rule for the 2D random walk is the same as 1D random walk in the
energy space.

\begin{figure}[h]
\caption{ The histogram of the two-dimensional random walk in energy and
order-parameter space for the $3D$ EA spin-glass model. }
\label{fig:sg3d_hist_3D}

\includegraphics{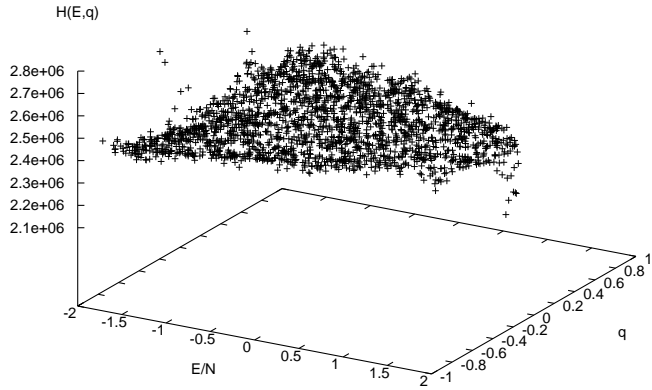}\vspace{3.5in}

\end{figure}

In Fig.~\ref{fig:sg3d_hist_3D}, we show the histogram of the 2D random walk
in energy-order parameter space, which is very flat. With the density of
states $G(E,q)$, we can calculate any quantities as we did in the previous
sections. It is very interesting to study the roughness of this model. First
we study the canonical distribution as a function of the order-parameter: 
\begin{equation}
P(q, T)={{\sum\limits_{E}G(E,q) e^{-E/k_{\text B}T} }}
\end{equation}

\begin{figure}[h]
\caption{ a) Overview of the rough topology of the canonical distribution in
the order-parameter space for one bond configuration of the $3D$ EA model on
an $L=6$ simple cubic lattice. b) The logarithmic plot for the canonical
distribution as a function of the order-parameter for the $3D$ EA model at
the temperature $T=0.5 $. }
\label{fig:sg3d_PQ_3D}

(a)\\
\includegraphics{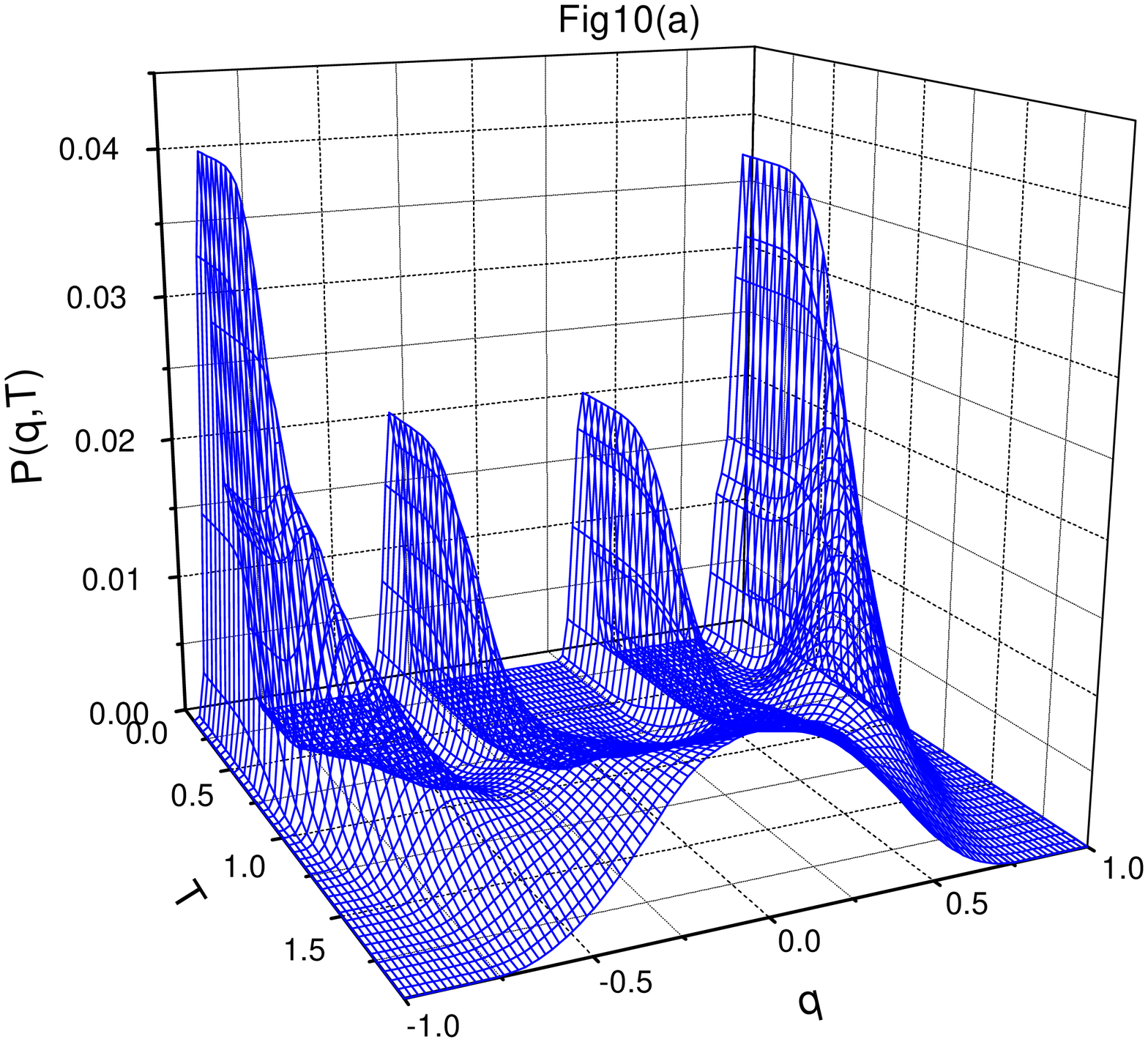}
\vspace{3.5in}

(b) \\
\includegraphics{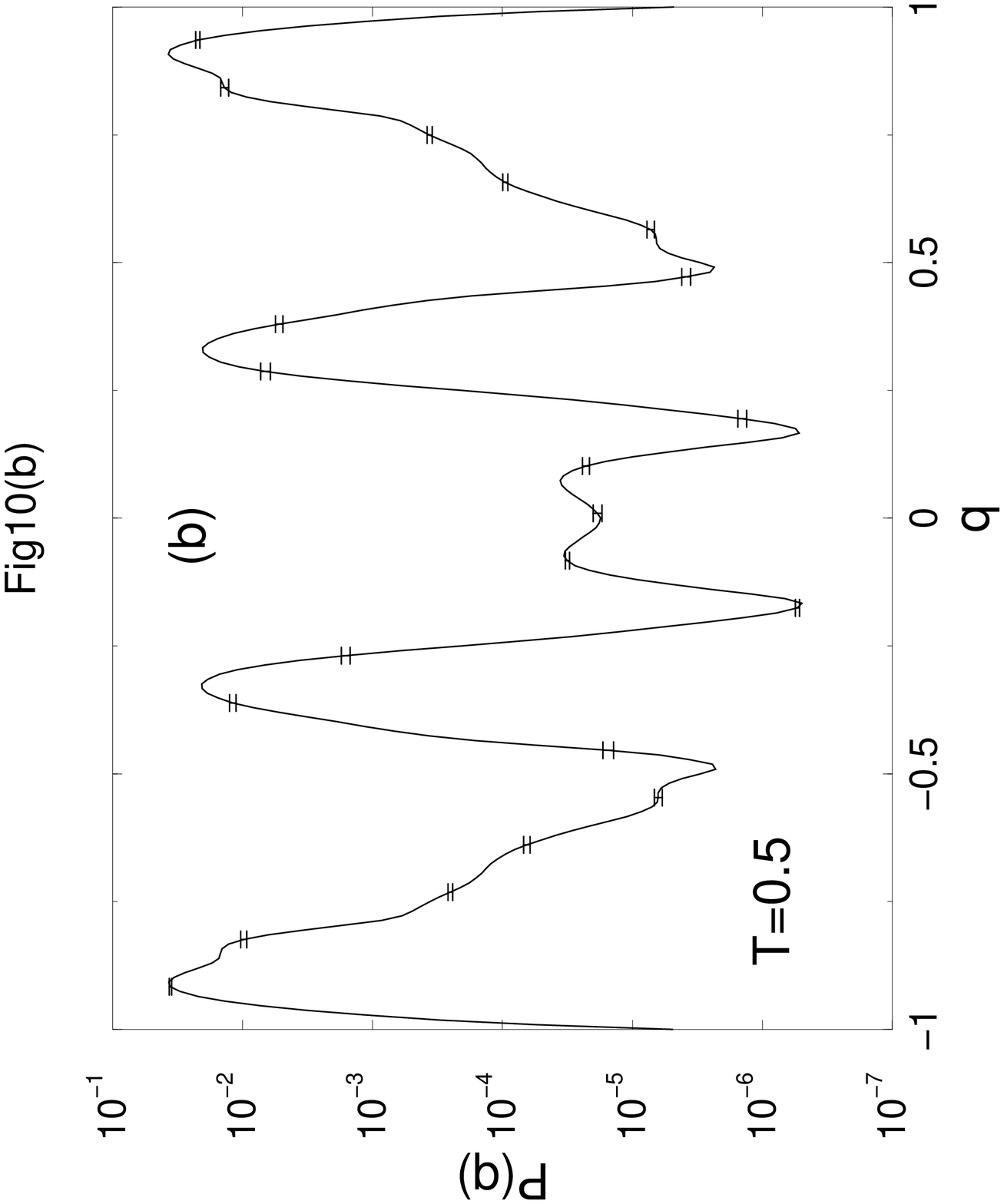}
\vspace{3.5in}

\end{figure}

In Fig.~\ref{fig:sg3d_PQ_3D}(a), we show a 3D plot for the canonical
distribution at different temperatures for one bond configuration of $L=6$
EA model. At low temperatures, there are four peaks, and the depth of the
valleys between peaks depends upon temperature. When the temperature is
high, the multiple peaks converge to a single central peak. Because we use
the linear scale to show our result in the Fig.~\ref{fig:sg3d_PQ_3D}(a). It
is not clear how deep the dips among peaks are. In Fig.~\ref{fig:sg3d_PQ_3D}%
(b), we show the canonical distribution using logarithmic scale for the same
distribution but only at $T=0.5$,  and we find that the dips are as deep as $%
10^{-4}$. We also noted there actually are six peaks,
but the plot with linear scale does not show all of them because two are  
as small as $10^{-3}$ compared to other four peaks.

\begin{figure}[h]
\caption{ (a) The canonical distribution $P(q)$ as a function of the
order-parameter for one bond configuration of the $3D$ EA model on an L=8
simple cubic lattice at the temperature $T=0.1\sim2.0$. (b) Corresponding
energy landscapes $U(q,T)$ at different temperatures. }
\label{fig:sg3d_L8_PQ}

(a) \\
\includegraphics{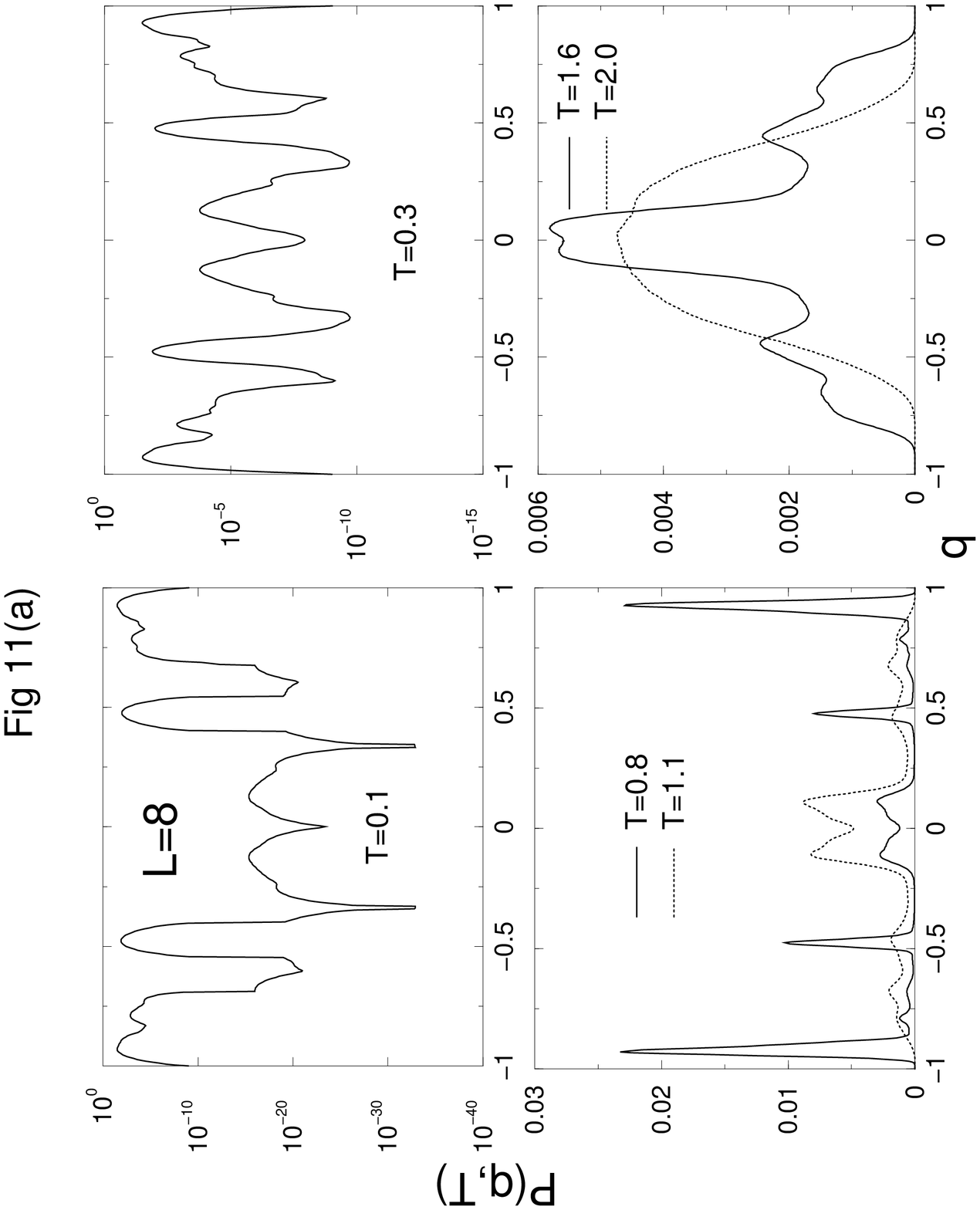}
\vspace{3.5in}

(b) \\
\includegraphics{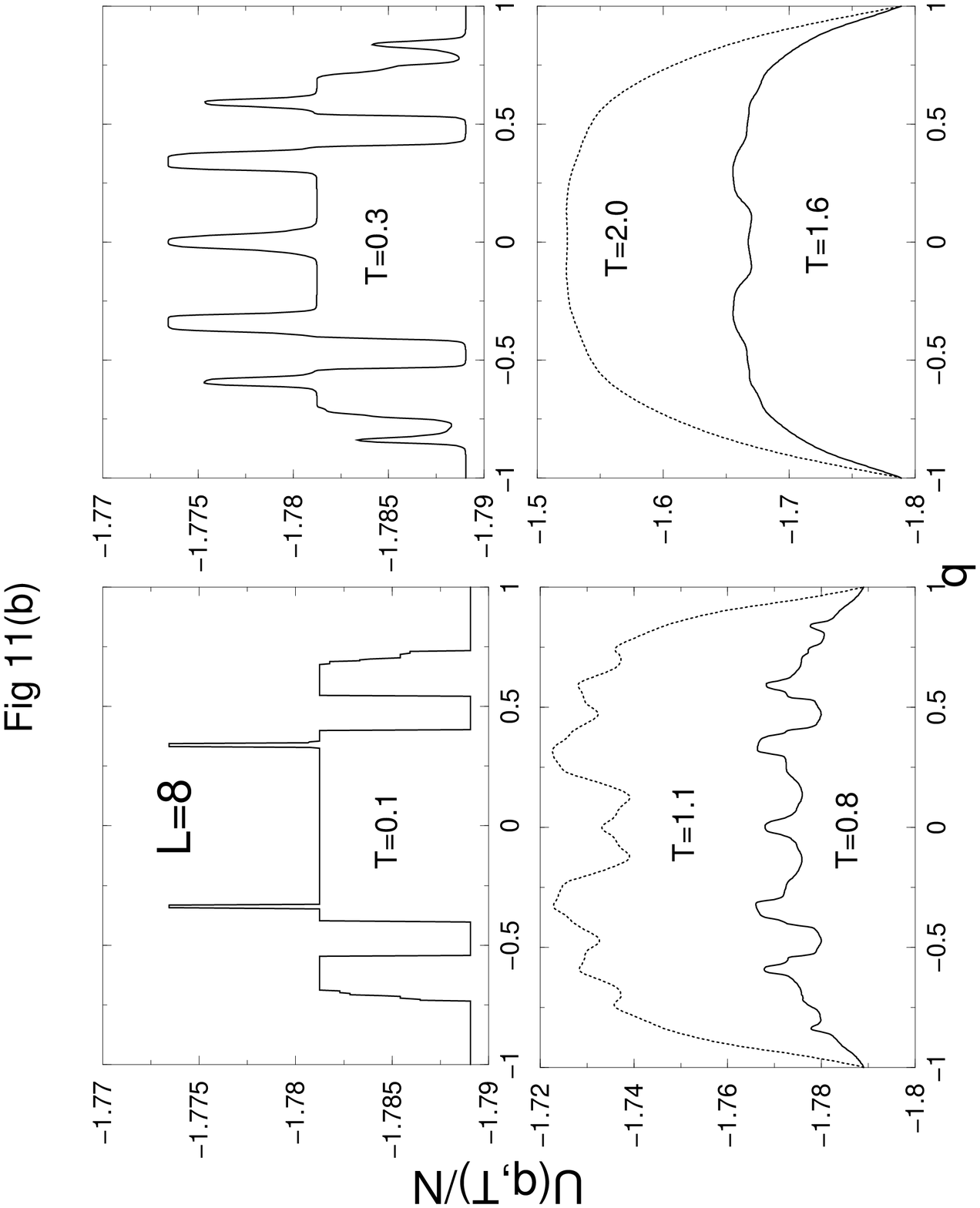}
\vspace{3.5in}

\end{figure}

In Fig.~\ref{fig:sg3d_L8_PQ}(a), we show the roughness of the canonical
distribution for another realization on an $8^3$ lattice. Because of the
wide variation in the distribution at low temperature we used a logarithmic
scale: the relative size of dips are as deep as $10^{-30}$ at $T=0.1$. There
are several local minima even at high temperatures. With conventional Monte
Carlo simulations, it is almost impossible to overcome the barriers at the
low temperature, so the simulation will get trapped in one of the local
minima as shown in the figure. With our algorithm all states will be visited
with more or less the same probability and trapping is not a problem.

With the density of states $G(E,q)$, we also can calculate the energy
landscape by: 
\begin{equation}
U(q,T)={\frac{{\sum\limits_{E,q} EG(E,q) e^{-\beta E} } }{{\sum\limits_{E,q}
G(E,q) e^{-\beta E} }}}
\end{equation}
In Fig.~\ref{fig:sg3d_L8_PQ}(b), we show the internal energy as a function
of order-parameter for temperatures $T=0.1 \sim 2.0$. We find that the
landscape is a very rough at low temperatures. The roughness of the energy
landscape agrees with the one for canonical distribution. But the maxima in
energy landscape are corresponding to the minima approximately in the
canonical distribution.

As we already noted in the previous paragraph, the roughness of the
landscape of the spin glass model makes the conventional Monte Carlo
simulation extremely difficult to apply. Therefore, even a quarter of a
century after
the model was proposed, we even can not conclude whether there is a
finite phase transition between glass phase and disordered phase. 
With Monte Carlo simulations on a  large system ($64^{2} \times 128$) and a
finite-size scaling analysis on a  small lattice, Marinari {\it et al.}~%
\cite{Marinari} expressed doubt about the existence of the
``well-established" finite-temperature phase transition of the 3D Ising spin
glass~\cite{Ogielski1,Bhatt1}. Their simulational data can be described
equally well by a finite-temperature transition or by a $T=0$ singularity of
an unusual type. Kawashima and Young's simulational data could also
not rule out the possibility of $T_{
{\text g}}=0$~\cite{Kawashima}. Thus even the existence of the
finite-temperature phase transition is still controversial, and thus the
nature of the spin glass state is uncertain. Although the best available
computer simulation results \cite{berg_2,Marinari_1,marinari_2} have been
interpreted as a mean-field like behavior with replica-symmetry
breaking(RSB)~\cite{parisi_0}, Moore {\it et al.} showed  evidence for
the droplet picture~\cite{fisher_huse} of spin glasses within the
Migdal-Kadanoff approximation. They argued that the failure to see droplet
model behavior in Monte Carlo simulations was due to the fact that all
existing simulations are done at temperatures too close to transition
temperature so that system sizes larger than the correlation length were not
used. As discussed in the previous paragraph, the lower the temperature is,
the rougher the canonical distribution and energy landscape are; hence, it
is almost impossible for conventional Monte Carlo methods to overcome the
barrier between local minima and globe minima. It is possible to heat the
system up to increase the possibility of escape from local minima by
simulated annealing and the more recent simulated tempering method~\cite%
{marinari} and parallel tempering method~\cite{hukushima,Hukushima_1}, but
it is still very difficult to perform equilibrium simulations at low
temperatures. Very recently Hatano and Gubernatis proposed a bivariate
multicanonical Monte Carlo method for the 3D $\pm J$ spin glass model, and
their result also favors the droplet picture~\cite{hatano,hatano_2}.
Marinari, Parisi {\it et al.} argued, however, that the data were not
thermalized~\cite{marinari_2}. The nature of spin glasses thus remains
controversial~\cite{Palassini_3}.

\begin{figure}[h]
\caption{ Properties of the $3D$ EA spin glass model calculated from the
density of states $G(E,q)$ resulting from a $2D$ random walk in energy -
order parameter space: (a) The order parameter vs temperature; (b) The
temperature dependence of the fourth order cumulant of the order parameter.
The cumulants for different lattice sizes cross around $T_{\text c}=1.2$. }
\label{fig:sg3d_QT}

(a) \\
\includegraphics{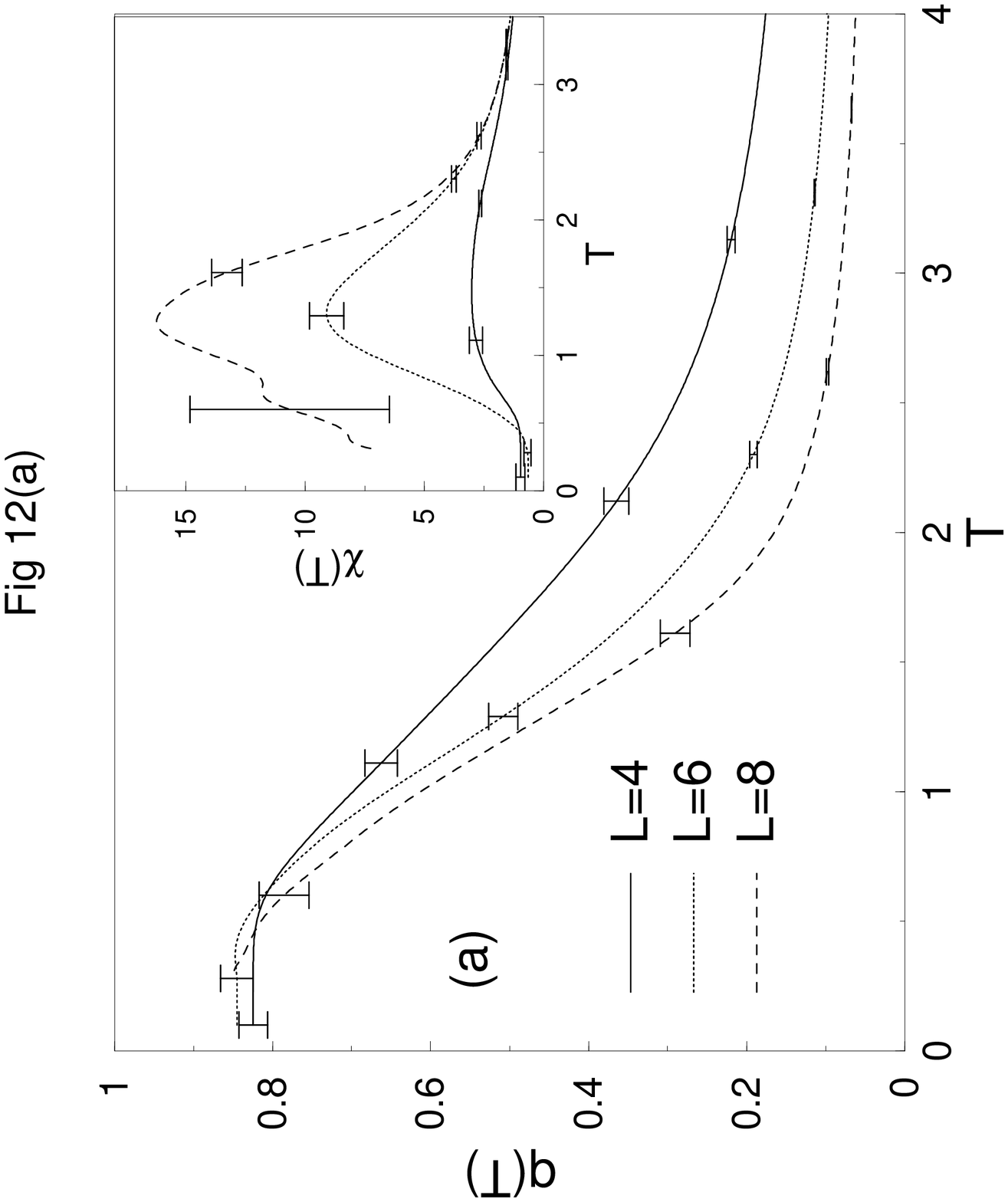}
\vspace{3.5in}

(b) \\ 
\includegraphics{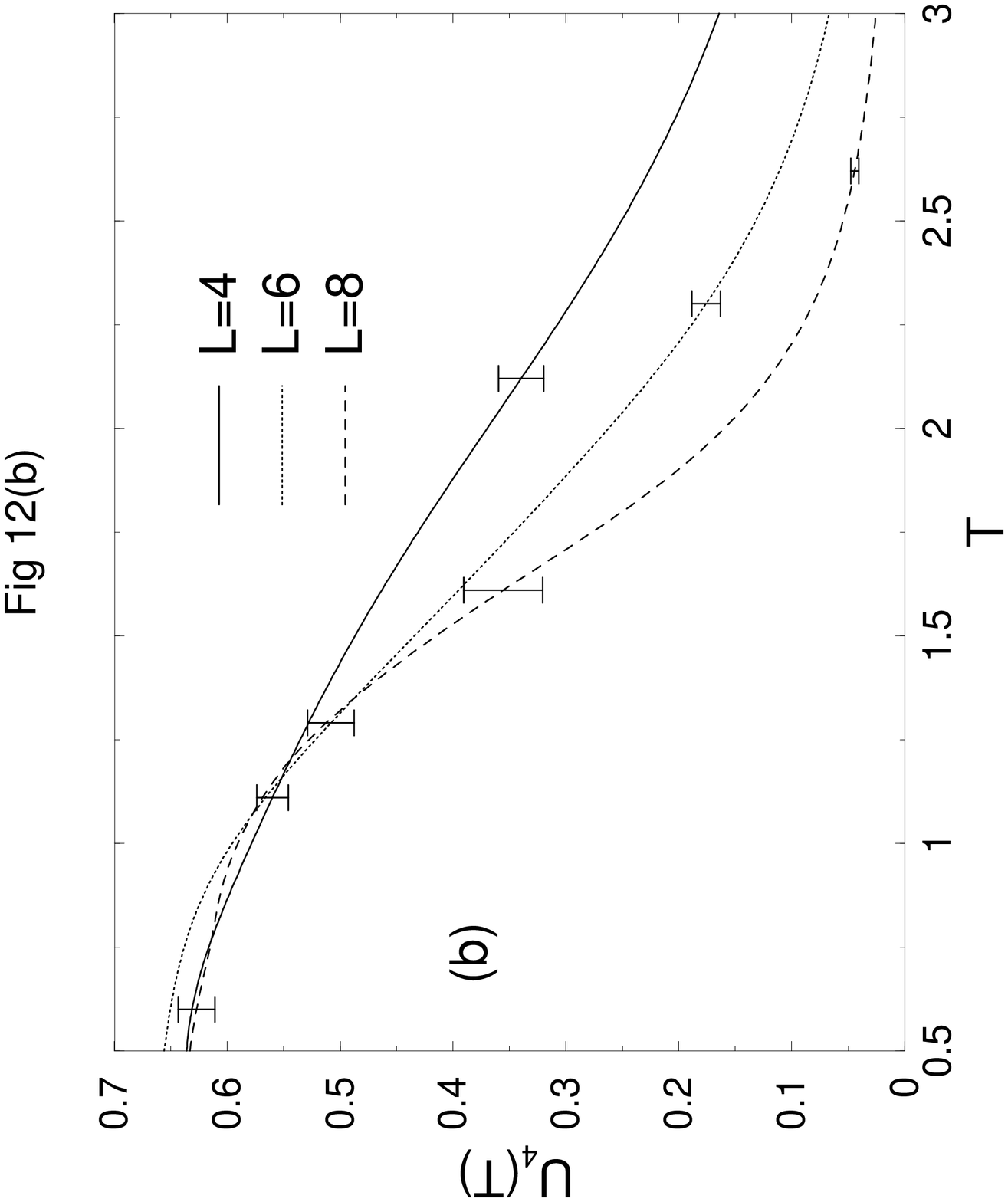}
\vspace{3.5in}

\end{figure}

The algorithm proposed in this paper provides an alternative for the study
of complex systems. Because we need to calculate the order-parameter with
high accuracy, and this quantity is not directly related to the energy, we
need to perform a random walk in the two-dimensional energy - order
parameter space. After we estimate the density of states in this $2D$ space,
we can calculate the order-parameter at any temperature from the
canonical average. In ~\ref{fig:sg3d_QT}(a), we show our results for the $3D$
EA model for $L=4$, 6 and 8. Because we need to perform a 2D random walk
with a total of about $L^6$ states, the simulation is only a practical for a
small system($L\leq8$). The results in the figure are the average over 100
realizations for $L=4$, 50 realizations for $L=6$ and 20 for $L=8$.

We notice that the behavior of $\langle q(T) \rangle$ is very similar to the
magnetization (the order-parameter for the Ising model), but the finite
value at low temperature is not necessarily equal to 1 because of the high
degeneracy of the ground state for the spin glass model. The fluctuation of
the order-parameter at the different temperatures for $L=4$, 6 and 8 is
shown in the inset of the figure.

To estimate the transition temperature of the spin glass system, we
calculated the fourth order cumulant as a function of temperature. In Fig. %
\ref{fig:sg3d_QT}(b), we show our simulational results for $L=4$, 6 and 8.
All curves clearly cross around $T_{\text g}=1.2 $. 
Below this temperature, the spin
configurations are frozen into some disorder  ground states and the order
parameter assumes a finite value. Above this temperature 
$T_{\text g}$, the system
is in a disordered states and the order parameter vanishes.

One complication for simulation of such random systems is the determination
of the relative importance of the error due to the simulation algorithm and
the error due to the finite sampling of bond distributions. From Fig.~\ref%
{fig:sg3d_QT} (a) and (b), we can not tell what the origin of the error bars
is so we also performed multiple independent simulations for the same bond
configuration on a $L=6$ 3D EA model. We found that the statistical errors
for the order parameter and the fourth order cumulant from these simulations
were much smaller than the error bars shown in Fig.~\ref{fig:sg3d_QT} (a)
and (b) for all temperatures. We conclude that the error bars in the figure
arise almost completely from the randomness of the system.

The computational resources devoted here to the EA model were not immense.
All our simulations for one bond configuration ($L=4$, 6 and 8) were
performed within two days on (multiple) Linux machines ($200-800$MHz) in the
Center for Simulational Physics. This effort should thus be viewed as a
feasibility study, and substantially more effort would be required to
determine the nature of the spin glass phase or to estimate the transition
temperature with high accuracy. Nonetheless, we believe that these results
show the applicability of our method to systems with a rough landscape.
Because of the number of states is about $N^2$ for $2D$ random walks, such
calculations not only require huge memory during the simulation but also
substantial disk space to store the density of states for the later
calculation of thermodynamic quantities.

\section{Discussion and Conclusion}

In this paper, we proposed an efficient algorithm to calculate the density
of states directly for large systems. By modifying the estimate at each step
of the random walk in energy space and carefully controlling the
modification factor, we can determine the density of states very accurately.
Using the density of states, we can then calculate thermodynamic quantities
at essentially any temperature by applying simple statistical physics
formulas. An important advantage of this approach is that we can also
calculate the Gibbs free energy and entropy, quantities that are not
directly available from conventional Monte Carlo simulations.

We applied our method to the $2D$ $Q=10$ Potts model which demonstrates a
typical first-order phase transition. By estimating the density of states
with lattices as large as $200 \times 200$, we calculated the internal
energy, specific heat, free energy and entropy in a wide temperature region.
We found a typical first-order phase transition with a ``discontinuity'' for
the internal energy and entropy at $T_{\text{c}}$. The first derivative of
the free energy also shows such a discontinuity at $T_{\text{c}}$. The
transition temperature estimated from simulational data is consistent with
the exact solution.

We also applied our algorithm to the $2D$ Ising model, which shows a
second-order phase transition. The density of states obtained by the end of
our simulations was compared directly with the exact solution on $50 \times
50 $ lattice. The relative errors for most important energy levels are less
than 0.019\%. It was also possible to calculate the density of states for a $%
256 \times 256 $ lattice with a computational effort of $6.1 \times 10^6$
Monte Carlo sweeps. With the accurate density of states, we calculated the
internal energy, specific heat, Gibbs free energy and entropy. For all
temperatures between $T=0$ and $T=8$, the relative errors are smaller than $%
0.09\%$ for internal energy, $0.0008\%$ for free energy, $1.2\%$ for
entropy and $4.5\%$ for specific heat.

We should point out that our simulational results for $L=256$ are close to
the same accuracy as the specific heat of the $2D$ Ising model for $L=64$
with the transition matrix Monte Carlo re-weighting method with $2.5 \times
10^7$ MC sweeps~\cite{wang_1}. Our estimate of the entropy for the 2D Ising
model for the $L=256$ lattice is even more accurate than the results
obtained for the same quantity for $L=32$ with the broad histogram method
and flat histogram method~\cite{wang_0}, which needed about $10^7$ MC
sweeps. Our simulational effort for $L=256$ was $6.1 \times 10^6$ MC sweeps.

The algorithm was also applied with success to the $3D$ $\pm J$ EA spin
glass model for which we could determine the roughness of the energy
landscape and canonical distribution in order-parameter space. The internal
energy and entropy at zero temperature were estimated up to a lattice size $%
20^3$, and the transition temperature was estimated as about 
$T_{\text g}=1.2$.

In this paper, we only concentrated the random walk in energy space (and
order-parameter space); however, the idea is very general and we can apply
this algorithm to any parameters~\cite{berg_4}. The energy levels of the
models treated here are perfectly discrete and the total number of possible
energy levels is known before simulation, but in a general model such
information is not available. Since the histogram of the random walk with
our algorithm tends to be flat, it is very easy to probe all possible
energies and monitor the histogram entry at each energy level. For some
models where all possible energy levels can not be fitted in the computer
memory or the energy is continuous, e.g. the Heisenberg model, we may need
to discretize the energy levels. According our experience on discrete and
continuous models, if the total number of possible energies is around the
number of lattice sites $N$, the algorithm is very efficient for studying
both first- or second-order phase transitions.

In this paper, we only applied our algorithm to simple models, but since the
algorithm is very efficient even for large systems it should be very useful
in the studies of general, complex systems with rough landscapes. It is
clear, however, that more investigation is needed to better determine under
which circumstances our method offers substantial advantage over other
approaches and we wish to encourage the application of this approach to
other models.

\acknowledgments

We would like to thank S. P. Lewis,  H-B Schuttler, 
T. Neuhaus and A. H\"uller for comments and
suggestions, K. Binder, N. Hatano, P. M. C. de Oliveira and  C. K. Hu 
for helpful discussions. We
also thank M. Caplinger for support on technical matters and P. D. Beale for
providing his Mathematica program for the calculation of the exact density
of states for the $2D$ Ising model. The research project is supported by the
National Science Foundation under Grant No. DMR-0094422.


\newpage




\newpage 

\newpage 

\newpage


\newpage 

\newpage

\newpage 


\newpage

\newpage 

\newpage

\newpage 

\widetext

\begin{table}[tbp]
\caption{ Estimates of ``transition temperature" $T_{\text c}$ and positions
of double peaks $E_1^{max}$, $E_2^{max}$ for the $Q=10$ Potts model with our
method, Multicanonical (MUCA) ensemble[5] 
and 
Multibondic (MUBO) cluster algorithm [6]. 
$E_1^{max}$ and $E_2^{max}$ are the energy per lattice site
at the two peaks of canonical distribution at $T_{\text c}$. 
}
\begin{tabular}{c|ccc|ccc|ccc}
size &  & Our method &  &  & MUCA &  &  & MUBO &  \\ \hline
L & $T_{\text c}$ & $E_1^{max}$ & $E_2^{max}$ & $T_{\text c}$ & $E_1^{max}$
& $E_2^{max}$ & $T_{\text c}$ & $E_1^{max}$ & $E_2^{max}$ \\ \hline
12 & 0.70991 & 0.8402 & 1.7013 & 0.710540 & 0.806 & 1.688 & 0.7105402 & 0.833
& 1.72 \\ 
16 & 0.70653 & 0.8694 & 1.6967 & 0.706544 & 0.844 & 1.676 & 0.7065144 & 0.867
& 1.71 \\ 
20 & 0.70511 & 0.8925 & 1.6875 &  &  &  & 0.7047891 & 0.883 & 1.69 \\ 
24 & 0.70362 & 0.8940 & 1.6765 & 0.703730 & 0.908 & 1.698 &  &  &  \\ 
26 & 0.70317 & 0.9002 & 1.6805 &  &  &  & 0.7034120 & 0.908 & 1.682 \\ 
30 & 0.70289 & 0.9233 & 1.6888 &  &  &  &  &  &  \\ 
34 & 0.70258 & 0.9343 & 1.6732 & 0.702553 & 0.927 & 1.683 & 0.7025530 & 0.921
& 1.676 \\ 
40 & 0.70239 & 0.9337 & 1.6731 &  &  &  &  &  &  \\ 
50 & 0.70177 & 0.9416 & 1.6776 &  &  &  & 0.7018765 & 0.940 & 1.674 \\ 
60 & 0.70171 & 0.9522 & 1.6733 &  &  &  &  &  &  \\ 
70 & 0.70153 & 0.9519 & 1.6717 & 0.701562 & 0.9511 & 1.670 &  &  &  \\ 
80 & 0.70143 & 0.9576 & 1.6701 &  &  &  &  &  &  \\ 
90 & 0.70141 & 0.9551 & 1.6727 &  &  &  &  &  &  \\ 
100 & 0.70135 & 0.9615 & 1.6699 & 0.701378 & 0.9594 & 1.6699 &  &  &  \\ 
120 & 0.70131 & 0.9803 & 1.6543 &  &  &  &  &  &  \\ 
150 & 0.70127 & 0.9674 & 1.6738 &  &  &  &  &  &  \\ 
200 & 0.70124 & 0.9647 & 1.6710 &  &  &  &  &  & \\
\hline
$\infty$   &  $0.701236\pm 0.000025$   &       &       & & & & & & \\ 
\hline
{\text {exact}}   &  0.701232...    &       &       & & & &  & & 

\end{tabular}%
\end{table}

\begin{table}[tbp]
\caption{ Estimates of entropy ($s_0$) and internal energy ($e_{0}$) per
lattice site at zero temperature for the 3D EA model by our method,
multicanonical Method(MUCA) [17]. 
}
\begin{tabular}{c|cc|cc}
size & Our Method &  & MUCA &  \\ \hline
L & $s_{0}$ & $e_{0}$ & $s_{0}$ & $e_{0}$ \\ \hline
4 & $0.075 \pm 0.027$ & $-1.734 \pm 0.006$ & $0.0724\pm0.0047$ & $%
-1.7403\pm0.0114$ \\ 
6 & $0.061 \pm 0.025$ & $-1.767 \pm 0.024$ & $0.0489\pm0.0049$ & $%
-1.7741\pm0.0074$ \\ 
8 & $0.0493 \pm0.0069$ & $-1.779 \pm 0.016$ & $0.0459\pm0.0030$ & $%
-1.7822\pm0.0081$ \\ 
12 & $0.0534 \pm 0.0012$ & $-1.780 \pm 0.012$ & $0.0491\pm0.0023 $ & $%
-1.7843\pm0.0030 $ \\ 
16 & $0.0575 \pm 0.0037$ & $-1.7758 \pm 0.0041$ &  &  \\ 
20 & $0.0556 \pm 0.0034$ & $-1.7745 \pm 0.0043$ &  & 
\end{tabular}%
\end{table}


\narrowtext

\end{document}